\documentclass[a4paper,12pt]{amsart}
\usepackage{amssymb,amsmath,mathrsfs}
\usepackage[usenames,dvipsnames]{xcolor}
\usepackage{hyperref}
\usepackage{tensor}
\usepackage{graphicx}
\usepackage[left=1in,top=1in,right=1in,bottom=1in,headheight=0in,foot=0.4in]{geometry}
\usepackage[nodisplayskipstretch]{sets pace} 
\usepackage{mdframed}
\usepackage{enumitem}
\usepackage[greek.polutoniko,english]{babel}
\usepackage{appendix}
\usepackage{tikz}

\usepackage{placeins}

\usepackage{float} % For precise float control

\expandafter\def\expandafter\normalsize\expandafter{%
    \normalsize%
    \setlength\abovedisplayskip{8pt}%
    \setlength\belowdisplayskip{8pt}%
    \setlength\abovedisplayshortskip{8pt}%
    \setlength\belowdisplayshortskip{8pt}%
}

\renewcommand{\footnoterule}{
  \kern 8pt
  \hrule width 0.4\linewidth height 0.4pt
  \kern 8pt % Change this value to increase or decrease the space
}

\makeatletter
\let\@email\email
\renewcommand{\email}[1]{\textbf{Principal Author:} \href{mailto:#1}{#1}}
\makeatother

\hypersetup{
	colorlinks=true,         
	linkcolor=MidnightBlue,          
	citecolor=MidnightBlue,
	urlcolor=MidnightBlue            
 }
 \let\oldmarginpar\marginpar
\renewcommand\marginpar[1]{\oldmarginpar{\color{red}\raggedright\scriptsize #1}}
\let\oldmarginpar\marginpar
\renewcommand\marginpar[1]{\oldmarginpar{\color{red}\raggedright\scriptsize #1}}

\setlist[enumerate,1]{label=\Roman*}
\setlist[enumerate,2]{label=\roman*}

\theoremstyle{definition}

\usepackage{natbib}
\setcitestyle{aysep={}} % author date
\usepackage{amsaddr}

\title[In Search of Cosmic Time]{\Large{In Search of Cosmic Time:\\ Complete Observables and the Clock Hypothesis}}

\author{Nicola Bamonti \textsuperscript{*}}
\address{\vspace{-0.8pc}\tiny{Department of Philosophy, Scuola Normale Superiore, Piazza dei Cavalieri, 7, Pisa, 56126, Italy}}
\address{\vspace{-0.8pc}\tiny{Department of Philosophy, University of Geneva, 5 rue de Candolle, 1211 Geneva 4, Switzerland}}

\author{Karim P. Y. Th\'ebault}
\address{\vspace{-0.8pc} \tiny{Department of Philosophy, University of Bristol, Cotham House, 29 Cotham Hill, Bristol BS6 6JL, UK}}

\thanks{\textsuperscript{*}\textit{Principal and corresponding author}: \href{mailto:nicola.bamonti@sns.it}{nicola.bamonti@sns.it}}

\begin{document}

\maketitle

\begin{abstract}
\singlespacing
This paper considers a new and deeply challenging face of the problem of time in the context of cosmology drawing on the work of \cite{Thiemann-k-essence,Thiemann:2007}. Thiemann argues for a radical response to the cosmic problem of time that requires us to modify the classical Friedmann equations. By contrast, we offer a conservative proposal for solution of the problem by bringing together ideas from the contemporary literature regarding reference frames \citep{Bamonti2023,BamontiGomes2024}, complete observables  \citep{Gryb:2016a,Gryb:2023}, and the model-based account of time measurement \citep{tal:2016}. On our approach, we must reinterpret our criteria of observability in light of the clock hypothesis and the model-based account of measurement in order to preserve the Friedmann equations as the dynamical equations for the universe.

\end{abstract}

\tableofcontents
\newpage
\setstretch{1.2}
\section{Introduction}

\subsection{Pr\'ecis}
\begin{itemize}[leftmargin=0cm]
\item [] \textit{A Cosmic Problem}: In every causally stable spacetime of general relativity there exists a symmetry invariant cosmic time function.  However, it is unclear whether we should treat such cosmic times as `observables' according to the standard `Dirac criterion' of observability deployed by scientists in analysing the theory.
\item [] \textit{A Radical Proposal}: Taking the problem of the observability of cosmic time seriously in the context of the Friedmann equations for cosmology leads to potentially significant modifications to the equations. In particular, one can show that consistently applying the standard Dirac criterion may modify the form of the equations in a manner that can have empirical consequences. 
\item [] \textit{Our Alternative Approach}: A different option is to reinterpret the criteria of observability in light of the clock hypothesis. On the assumption that there are clocks which measure proper time along any given world-line, one is able to treat cosmic time as an observable and preserve the form of the Friedmann equations.
\item [] \textit{The Conservative's Dilemma}: Whatever option we take, something must change, for things to stay as they are.
\end{itemize}

\subsection{The Problem of Cosmic Time}
Spacetime symmetry and time evolution are not straightforward to reconcile in the context of the diffeomorphism invariance of general relativity. In particular, the ability to `re-slice' relativistic spacetimes into arbitrary decompositions of space-like hypersurfaces indicates that time is `many-fingered' within the theory.\footnote{The implications of diffeomorphism invariance and the related notions of general covariance and background independence are much debated. See \cite{pooley:2017,James_Read2023-mk}. Many-fingered time and the decomposition into space-like hyperspaces requires relativistic spacetimes to be globally hyperbolic \citep{Geroch1970,misner2017gravitation}.} In the context of cosmology, however, it is standard to make reference to a notion of `cosmic time'. One application of cosmic time is, of course, in discussions of the age of the universe. Most vividly, in simple FLRW cosmological models the cosmic time function labels isotropic and homogenous spatial hypersurfaces \textit{and} corresponds to the temporal length of geodesic paths that run from the big bang until the present day.\footnote{An important connected notion of time in general relativity is `York time'. This corresponds to the time that labels spatial slices of constant mean curvature and has important connections with conformal invariant approaches to the initial value problem of general relativity \citep{york:1972,york:1973,gomes:2011,gryb:2016c}. York time corresponds to cosmic time for homogenous spatial slices, meaning that in FLRW models the constant mean curvature slices are the same as the slices of constant cosmic time. However, in general, the two are distinct (see \cite{Roser2014}).} 
Cosmic times can be shown to be well-defined in any relativistic cosmology with stable causal structure \citep{hawking:1969} and \textit{are} consistent with the spacetime symmetries of general relativity since they can be defined in a manifestly spacetime diffeomorphism invariant manner.\footnote{Formally, a cosmic time function on a spacetime manifold $M$ is a function that assigns to any $p \in M$ the supremum of the durations of all future-directed continuous timelike curves ending at $p$ \citep{andersson:1998,Fletcher}. See \cite{smeenk:2013,callender:2021} for philosophical overviews of time in cosmology. See \cite{rugh:2009} for a discussion focused on cosmic time.} 

There are, however, plausible formal and physical reasons to doubt the status of a cosmic time as an \textit{independent physical observable}. First, on very general grounds, we might demand of an observable that it corresponds to a measurable quantity of the theory and it is not immediately clear how one would measure cosmic time since there is no natural `clock' that we would understand to read off cosmic time nor clear methodology to construct one. Second, we might also demand that genuine observables are \textit{relational} in the sense that they correspond to \textit{correlations} between measurable quantities. On such an understanding of what is to be an observable we would not expect that cosmic time is observable since it is does not appear to have the right form to be a correlation between two measurable quantities. 

This potential challenge to the observable status of cosmic time becomes all the more acute in the context of the canonical formalism for general relativity. A remarkable achievement of the canonical gravitational formalism, pioneered by Dirac and ADM \citep{Dirac:1958b,ADMII,ADMReview}, was to show that one can represent spacetime symmetries within a 3D+1 space and time formalism in \textit{constrained Hamiltonian terms}. In this formulation, the spacetime diffeomorphism symmetry of the 4D covariant theory is encoded in the \textit{hypersurface deformation algebroid of constraints}. This algebroid is a mathematical structure encoding how the constraints governing the dynamics of the 3D spatial slices transform under infinitesimal normal and tangential deformations of the slices \citep{Teitelboim:1973,Gryb2016b}. These deformations correspond to `shifts' in time and space and ensure that the canonical Dirac formalism preserves the underlying covariance of general relativity, even within the 3D+1 decomposition. Although it is restricted to spacetimes admitting globally hyperbolic topology, there is no expectation that such a limitation will conflict with the application of the formalism to cosmology (the black hole case is more subtle due to the Kerr solutions not being globally hyperbolic).\footnote{The general expectation that all physically realistic spacetimes must be globally hyperbolic is based upon assumptions regarding the \textit{instability} of features such as Cauchy horizons that lead to a breakdown in global hyperbolicity. See \citealp[\S12]{Penrose:1980ge}, \cite[p. 202]{Wald1984}.} A further attractive feature of the canonical formulation of general relativity is that it comes equipped with a formal criterion, known as the \textit{Dirac criterion}, which identifies a quantity as an observable if it has (weakly) vanishing Poisson bracket with all first-class constraints of the theory. Furthermore, in a groundbreaking and highly influential analysis, it was shown that a particular approach to relational conception of an observable, as a correlation between two \textit{partial observables} leading to a \textit{complete observable} \citep{Rovelli_2002}, provides a construction of objects that satisfy the Dirac criterion. An unambiguous criterion for functions to be observables is of course indispensable staging material needed towards the pursuit of canonical quantisation. It is the algebra of classical observables that one seeks to faithfully represent as operators on a physical Hilbert space. Such a criterion is also useful for settling questions of observability in the classical formalism. The problem, however, is that it is not at all clear that canonical representations of cosmic time would in fact satisfy the Dirac criterion. In particular, prima facie, cosmic time does not appear to be the right kind of relational object to fit with the complete observables approach. It is not at all clear that cosmic time is either a partial or complete observable in Rovelli's terms. This is, ultimately, a cosmological manifestation of the infamous problem of time, which we name \textit{the cosmic problem of time} and upon which we will have more to say later.

The acuteness of the cosmic problem of time can be illustrated most clearly in the context of the Friedmann equations that are the cornerstone of the standard model of cosmology. These equations describe the expansion of the universe in terms of the dynamics of the \textit{spatial scale factor}, $a$, with respect to the \textit{cosmic time parameter}, $t$. These equations can be straightforwardly represented in canonical terms since they turn out to simply be equivalent to the Hamilton equations. In this case it is simple to observe that \textit{neither the cosmic time nor the scale factor are Dirac observables}. As per our discussion above, by the Dirac criterion, such observables must be phase space functions which have (weakly) vanishing Poisson bracket with the Hamiltonian constraint that generates the dynamics of the Friedmann equation. However, clearly neither $a$ nor $t$ can satisfy such a criterion. This is because, in the first case, $a$ evidently has non-zero Poisson bracket with the Hamiltonian for any non-trivial cosmological dynamics by the definition of the Friedmann equations. In the second case, not only is $t$ not naturally understood as a phase space variable, rather it is a parameter of a phase space flow, but any phase space function that `marches in step' with $t$ will have non-zero Lie drag with the Hamilton vector field of the Hamiltonian, and thus non-vanishing Poisson bracket. Though simple to articulate and difficult to ignore once it is recognised, this problem has received almost no discussion in physics or philosophy. 

Isolation of the problem crucially depends upon how strict one is regarding the application of the Dirac criterion in the context of `partial observables' -- both in general and in the cosmological case. Important discussions of this issue in general and in the cosmological context can be found in the work of the noted theoretical physicist Thomas Thiemann. The general question of observability of partial observables is discussed in his textbook \cite{Thiemann:2007}. There he argues that `a measurable quantity is always a complete observable, even pointers of a clock are observables and not partial observables. Now complete observables are defined with respect to non-measurable quantities...which we will simply call non-observables' (p. 78). Thiemann's view implies that both the scale factor and the cosmic time parameter are \textit{non-observables} and that the Friedman equations therefore \textit{do not} describe the evolution of observable quantities.  Indeed, an analysis along precisely these lines can be found in  \cite{Thiemann-k-essence}. There Thiemann explicitly argues that `it is incorrect to interpret the FLRW equations as evolution equations of observable quantities' (p. 9). Rather, he suggests, we should follow the complete observables procedure and re-write the equations relationally. The true evolution equations can then shown for a simple model to acquire \textit{empirical modifications} when compared to the Friedmann equations. We are thus lead from a formal-conceptual problem regarding time and observability to a physical proposal for a new approach to cosmology with empirical consequences. One of the major goals of this paper is to consider this chain of argument leading to Thiemann's radical cosmological revisionism.

The second major goal of this paper is to formulate and defend a conservative approach to finding cosmic time based upon the application of the \textit{clock hypothesis} in the context of the concept of \textit{Hubble flow}.\footnote{The idea of \lq{}operationalising\rq{} cosmic time via Hubble flow is a well-known standard approach in the literature on FLRW cosmology. See e.g. \cite[\S\ 27.4]{misner2017gravitation} or also \cite{Ryden_2016}.} The development of this alternative approach will lead us, in turn, to reconsider the methodological status of the clock hypothesis in cosmology and the meaning of `measurability' in the context of the complete observables programme. Our analysis draws crucially on the different types of reference frames in physical theory and the insights that this can deliver for clarifying foundational questions regarding the construction of `complete observables' as correlations between `partial observables'. Our goal is to synthesise key ideas from the contemporary literature regarding reference frames \citep{Bamonti2023,BamontiGomes2024}, complete observables  \citep{Gryb:2016a,Gryb:2023}, and the model-based account of time measurement \citep{tal:2016}, whilst drawing attention to a new and deeply challenging face of the problem of time in the context of cosmology. The cosmic problem of time leads to a dilemma: we can apply a conservative understanding of Dirac observables, downplay the significance of the clock hypothesis, and modify the Friedmann equations; or we can reinterpret our criteria of observability in light of the clock hypothesis and the model-based account of measurement, and preserve the Friedmann equations. Whatever option we take, something must change, for things to stay as they are. 

\subsection{Roadmap}
Section \ref{COandRRF} provides a brief overview of the complete and partial observables programme as a response to the problem of time. We then apply recent work on reference frames to disambiguate two important details in the definition of a partial observable and better understand what it means for a physical variable to play the role of a clock in the context of a complete observable. Section \ref{TinFE} considers the status of time in the Friedmann equations, poses Thiemann's challenge to the standard interpretation of these equations as dynamical equations and reviews his solution in terms of a de-parametrisation via a \textit{phantom field} through the well-known \cite{Brown1995} mechanism. Section \ref{HFandCH} reframes comic time as a proper time parameter $\tau$ along the Hubble flow. In this context, we consider the question of the observability and measurability of Hubble parameter (i.e. $H=\dot{a}/a$) and the question of whether we can consider $H(\tau)$ to be a complete observable whose dynamics is described by the unmodified Friedman equations.  Problematically, there is no physical system which \textit{even approximately} measures proper time along the Hubble flow. Section \ref{sec5} articulates our proposed solution: the introduction of a more liberalised sense of `measurement procedure' in the context of cosmic time. This more liberalised notion draws upon the model-based account of time measurement developed in the context of atomic clocks by \cite{tal:2016}. 

\section{Complete Observables and Real Reference Frames}
\label{COandRRF}
The problem of time is best understood as a cluster of formal, physical and conceptual challenges to the isolation of the physical degrees of freedom in theories which display temporal diffeomorphism symmetry.\footnote{See \cite{Kuchar:1992,Isham:1992ms,anderson:2017} for scientific overview. \cite{casadio:2024} provides an overview of a family of alternative approaches in which first-class phase-space constraints may be relaxed based on an interpretation of them as fixing the values of new degrees of freedom. Technically informed discussion in the philosophical literature include \cite{belot_earman_2001,Belot:2007,thebault:2019}. A hybrid formal and philosophical monograph-length treatment of the global problem of time is \cite{Gryb:2023}. Further references will be given where relevant below.} In the canonical representation, many of the challenges stem from lack of an unambiguous  phase space representation of re-foliation symmetry and the implications that this has for quantisation. However, the problem is not restricted to canonical representations, and reoccurs in covariant form, for example, in terms of the challenge of finding an appropriate measure in path-integral approaches. One particular pressing aspect of the problem is the tension between the standard definition of a gauge-invariant observable and the seemingly obvious fact that observable quantities change. In particular, in the context of constrained Hamiltonian theories, following \cite{dirac:1950,Dirac:1958a,Dirac:1964} the criterion to be an observable is to have (weakly) vanishing Poisson bracket with first class (primary) constraints.\footnote{This idea also traces back to the discussions of  \cite{bergmann:1956,Bergmann:1960,Bergmann1961,Bergmann61b,Bergmann:1962} and so one might plausibly use the term Bergmann-Dirac observables. However, Bergmann changed his view at various points. See  \cite{pitts:2019} for discussion.} Even for simple theories, temporal diffeomorphism symmetry immediately leads to a problem of time since, in such theories, we have that the Hamiltonian is a sum of first class constraints. Application of the Dirac criterion then immediately implies that observables are condemned to be frozen as constants of the motion. In general relativity this problem recurs in a more complex fashion but with essentially the same elements. There is an infinite family of Hamiltonian constraints and if we insist that Dirac observables commute with them, then the observables of general relativity are frozen. 

The standard approach to the problem of reconciling the Dirac definition of observables with the necessity to describe dynamics is the \textit{partial and complete observables} approach. This approach was pioneered by \cite{rovelli1991a,Rovelli1991,Rovelli_2002,Rovelli2004,Rovelli:2007} and later formalised by  \cite{Dittrich2006,Dittrich2007}.\footnote{For detailed overview see \citep{Thiemann:2007,Tambornino2012}. Important developments of the approach include  \citep{gambini:2001,gambini:2009} and \citep{Bojowald:2011,bojowaldhohn:2011b,hohn:2019b}. Critical responses include \citep{Kuchar:1991,Kuchar:1992,kuchar:1999,dittrich:2017}. For a review of the various notions of observable, that includes discussion of the limitations of the partial and complete observables approach, see \citep{anderson:2014,anderson:2017}.
For analysis of relationship between the partial and complete observables approach and earlier influential work by \cite{page:1983} see the analysis of \cite{hohn:2021} -- this work demonstrates the equivalence between the two approaches by putting the latter on a more rigours footing and resolving problems within its presentation as highlighted by \cite{Kuchar:1992}. For philosophical analysis of the ontological implications of the partial and complete observables approach an excellent extended discussion can be found in \citep[pp. 161--171]{Rickles:2007}. For a further overview see \citep{thebault:2019}, which contains further references and discussion.} The essence of this approach is to designate a subset of measurable quantities or `partial observables' as internal clocks, and then use correlations between the remaining variables and these clocks to construct `complete observables' that are both predicable and measurable, and which correspond to Dirac observables. More specifically, the formal application of the approach requires one to consider, for each Hamiltonian constraint, one physical variable to play the role of a physical clock. One first constructs parametrised flow expressions for the `evolution' of the clock and non-clock variables under the Lie flow of the vector field associated with the constraint; the simplest way to do this is via the relevant Hamilton-Jacobi equation \citep{Rovelli2004,Gryb:2016a,Gryb:2023}. One next inverts the flow equation for the clock variable and substitutes it into flow equations for non-clock variables to construct an algebraic expression for their correlation that is parameter free. Finally, one considers the correlation between the clock variable and the other variables at a particular value of the clock variable. This is a complete observable and corresponds to the value of the non-clock partial observables when the clock partial observables takes a particular value.

Let us provide the simplest possible physical example so the reader can conceptualise clearly how the procedure works. Consider two free particles moving in one dimension and described by a theory with a single Hamiltonian constraint. We can write an expression for the integral of motion in terms of the time parameter of the flow of the vector field associated with constraint, by solving the Hamilton-Jacobi equation. This will give us an expression for the position of each particle, \( q_i \), as a functions of the constants of motion (i.e. initial position and momenta), \( Q_i \) and \( P_i \), and parameter time, \( t \). This takes the form:
\begin{equation}\label{qflow}
q_i(t) = Q_i + \frac{P_i}{m_i} t 
\end{equation}
for $i=1,2$.  These variables are partial observables and do not commute with the Hamiltonian constraint $\mathcal{H}$, since we have that $\{\mathcal{H},q_i\}=\dot{q}_i \neq 0$. However, we can combine the two expressions for for $i=1,2$ to describe the correlation between the values of the position of each particle. We do this by inverting the expression for one variable such that we obtain $t$ as a function of $(q_i,Q_i,P_i)$, and then inserting this expression into the expression for the other. The first variable is then playing the role of a physical clock and we evaluate the second variable for a given value of the second variable, say $s\in\mathbb{R}$. In this way we get a family of complete observables, one for each value of $s$.\footnote{Note that this is also the construction of the so-called evolving constant of motions \citep{rovelli1991a}. In fact, a complete observable formally coincides with an evolving constant. The difference between the two concepts lies mainly in the fact that for evolving constants, the focus is on the evolution of the quantity with respect to the parameter $s$ that serves as ‘internal time’.} For even slightly complicated physical systems the inversion step may run into significant obstacles and is typically such that we can only define the relevant expressions for restricted values of the time parameter. \cite{Dittrich2007} provides a detailed treatment of such a case. In our case, by contrast, since the physical dynamics is trivial  and we are able to solve Hamilton equations for the considered system, the inversion is simply given by:
\begin{equation}
t = \frac{m_1}{P_1} (q_1 - Q_1) \label{tequation}
\end{equation}
Re-inserting this into \eqref{qflow}, we get:
\begin{equation}
q_2(q_1) = Q_2 - \frac{P_2}{m_2} \frac{m_1}{P_1} (q_1 - Q_1)
\end{equation}
Finally, we evaluate our expression $q_2(q_1)$ at $q_1=s$ to get the parametrised family of complete observables:
\begin{equation}
q_2(q_1)|_{q_1=s} = Q_2 - \frac{P_2}{m_2} \frac{m_1}{P_1} (q_1 - Q_1)|_{q_1=s}
\end{equation}
This is a complete observable constructed according to the Rovelli-Dittrich procedure. It is also a Dirac observable since for any
specification of $s$ we have $q_2(q_1)|_{q_1=s} : \Gamma \rightarrow \mathbb{R}$ and $\{\mathcal{H}, q_2(q_1)|_{q_1=s} \} = 0$, where $\Gamma$ is four dimensional phase space $(q_i,p_i)\in\Gamma$ for $i=1,2$.

There is a specific tension within the physics literature regarding the interpretation of the partial observables. This tension will prove crucial in the context of application of the approach to cosmology. Consider, in particular, that according to the original approach of \cite{Rovelli_2002}, by definition, a partial observable is `a physical quantity with which we can associate a (measuring) procedure leading to a number' (p. 2).\footnote{In his original definition Rovelli makes clear that the definition should not be understood in operationalist terms. In particular, he notes `The operational tone of the [partial observable] definition does not imply any adherence to operationalism here \citep{bridgman:1927}: the reference to measuring procedures is just instrumental for clarifying a distinction.' (ibid. p. 2). Crucially, the partial observables are well-defined theoretical quantities whose \textit{definition} does not require specification of a measurement procedure. For more discussion of operationalism see \cite{sep-operationalism,fankhauser:2021}.}  By contrast, following \cite{Thiemann:2007} we have that `a measurable quantity is always a complete observable, even pointers of a clock are observables and not partial observables. Now complete observables are defined with respect to non-measurable quantities...which we will simply call non-observables' (p. 78). A third view is advocated by  \cite{Gryb:2016a,Gryb:2023} in the context of theories with a single Hamiltonian constraint. On this approach, one can think of the complete and partial observables programme as allowing us de-parametrise  evolution purely in terms of observable quantities. However, this evolution is fundamentally controlled by the evolution equations generated by the Hamiltonian constraint and is \emph{always} well-defined, even when a particular deparametrisation breaks down. On this approach, even if one wishes to use parameter-free complete observable expressions, one is still required to retain the full partial observables representation. This supports the \cite{Rovelli_2002} perspective, in which partial observables are measurable quantities, rather than the \cite{Thiemann:2007} perspective, where the partial observables are understood as non-measurable.

A further important disambiguation can be made based upon the connection between partial and complete observables and reference frames. The role of reference frames in general relativity has an extensive philosophical literature.\footnote{See \cite{Earman1973,Earman1974}, \cite{Norton1989,Norton1993-NORGCA}, \cite{sep-spacetime-iframes}. Recently, a community of scholars has also emerged in the field of the so-called quantum reference frames, see \cite{Giacomini2021,QuantumHole} and reference therein.} Most relevant to our analysis is the distinction made by \cite{Bamonti2023} between:  `Idealised Reference Frames' (\textbf{IRF}s), in which any dynamical interaction of the material system represented by the reference frame is ignored; `Dynamical Reference Frames' (\textbf{DRF}s), in which the set of equations that determine the dynamics of the matter field is included but the the stress-energy tensor of the matter field used as reference frame is neglected; and `Real Reference Frames' (\textbf{RRF}s) in which both the dynamics of the chosen material system and its stress-energy tensor are taken into account. This distinction allows us to disambiguate two important details in the definition of a partial observable, which is often not stressed in the relevant literature.
Following \cite{BamontiGomes2024}, we should understand partial observables to be \textit{relational but gauge-variant} quantities that are nevertheless associated with a measuring procedure. We can understand this seeming contradiction in terms of the fact that partial observables are defined relative to an Idealised Reference Frame. In particular, the parameter of flow equation acts as an \textbf{IRF} and, as such, partial observables are relational in the sense that they describe the correlation between a physical variable and the \textbf{IRF}.  Consider our expression for the partial observables \eqref{qflow} above. The variables $q_i(t)$ are measurable quantities of the theory but they are not measurable independently of a specification of the value of the flow parameter $t$. Furthermore, since  \lq{}\lq{}all measurements are comparisons between different physical systems\rq{}\rq{} \cite[p.128]{Anderson1967-en},\footnote{See also \cite[p.99]{Rovelli2014}: \lq{}\lq{}In physics, when we talk about measurement, we refer to an interaction between a measured system $S$ and a measuring apparatus $O$.\rq{}\rq{}} $t$ itself represents a physical system, whose dynamics is neglected as a result of approximations. The second relevant remark is that not every pair of physical quantities to which measuring instruments can be associated can play the role of partial observables. \textit{Bona fide} partial observables must be dynamically coupled to each other, in order for their relation to constitute a \textit{bona fide} complete observable \citep{BamontiGomes2024}.

Let us then consider the status of the complete observables. In this context, the partial observable that is chosen as the clock observable is playing the role of a reference frame. The point above suggests that to construct a complete observable, we must use \textbf{DRFs} or \textbf{RRFs}. Since we are considering finite dimensional particle mechanics there is no stress-energy tensor to consider. However, the distinction between Dynamical Reference Frames and  Real Reference Frames can still be made. That is, a clock variable is \textit{always} a \textbf{DRF} since its dynamics is always relevant via the flow equation. However, it is only an \textbf{RRF} when the coupling between the clock variable and the other non-clock partial observables is included. Our simple system with free particles is thus an implementation of the complete observables programme in terms of a \textbf{DRF} rather than an \textbf{RRF}.\footnote{This connection also points to the sense in which the idea of inertial reference frames as discussed in the late nineteenth century by Lange, Neumann, Tait and others are examples of \textbf{DRF}s and not \textbf{RRF}s (see \citealp[pp.101-104]{barbour2000end} and \citealp[\S12]{barbour2001discovery}). This is one way of thinking about Mach's criticisms (see \citealp{thebault:2021b}).} However, in cases where the coupling is included, complete observables admit an interpretation in terms of an \textbf{RRF}. This is precisely the application of the complete observables approach that we will consider in the context of cosmology in the following section.   

\section{Time and the Friedmann Equations}
\label{TinFE}
The universe is estimated to be 13.7 billion years old. This estimation is made based upon the standard model of cosmology  --- the so-called $\Lambda CDM$ model --- in which the spacetime structure of the universe is described via general relativity with a cosmological constant $\Lambda$ \citep{weinberg1972gravitation}.\footnote{The \lq{}CDM\rq{} part stands for \lq{}Cold Dark Matter\rq{}, which is a form of matter that does not interact with electromagnetic radiation (hence \lq{}dark\rq{}) and moves slowly compared to the speed of light (hence \lq{}cold\rq{}). Cold dark matter is not universally regarded as being composed of material particles within an \lq{}extended Standard Model\rq{}. For a detailed discussion, see e.g. \cite{TURNER2000619}.} The base-model of modern cosmology, upon which more sophisticated models are built, is one in which the spatial structure is extremely simple. The \textit{Cosmological Principle} is defined by the condition that the universe is spatially homogeneous and isotropic on large scales ($\geq 10^2 MPc$). This means that, from any location, the distribution of matter and energy appears the same and without any preferred direction. 
FLRW spacetimes are the class of generally relativistic spacetimes that are spatially isotropic and homogenous and satisfy certain physically motivated energy conditions or equations of state.\footnote{The original papers are \cite{Friedman1922} \cite{Lematre1931}, \cite{Robertson1935}, \cite{Walker1937}. A formally precise overview is given in \cite[\S2.11]{Malament:2012}.} FLRW spacetimes are the formal basis for modelling the large-scale structure and dynamics of the universe with realistic models involving perturbations about an FLRW metric. 

In the context of the FLRW metric and a perfect fluid model of matter, the Einstein Field Equations take on the remarkably simple form given by the Friedmann equations. These equations describe the evolution of a single geometric variable, the scale factor $a(t)$, and take the form:
\begin{eqnarray}
\left( \frac{\dot{a}}{a} \right)^2 &=& \frac{8 \pi G}{3} \rho - \frac{kc^2}{a^2} + \frac{\Lambda c^2}{3} \\
\frac{\ddot{a}}{a} &=& -\frac{4 \pi G}{3} \left( \rho + \frac{3p}{c^2} \right) + \frac{\Lambda c^2}{3}
\end{eqnarray}
 where $t$ is the cosmological time, $\rho(t)$ and $p(t)$ are the density and pressure of the matter, $k=0,-1,+1$ is the spatial curvature parameter, and the constants have their usual meaning.\footnote{The value of $k$ does not fix the overall topology. In fact, different topological choices are possible for the same $k$: for example, a hyperplane (closed topology) is
characterised by curvature parameter $k= 0$, like a hyperplane (open topology).} For a given specification of matter-energy we can solve these equations to get dynamical expressions for the scale factor. For the matter-energy mix that we take to correspond to our universe (including dark energy and dark matter) the relevant expressions describe an expanding universe which matches our observational data to a remarkable degree -- although there are existing challenges \citep{Smeenk_2022,perivolaropoulos:2022,colgain:2024}. This model provides a standard, textbook level story of the expansion of the universe that is assumed by almost all cosmologists to be unproblematic, at least back to the (presumed) inflationary epoch. 

%Specifically, observations of the Cosmic Microwave Background (CMB), a relic radiation from the early Universe, provided strong support for the model’s predictions of a hot, dense origin characterising the $\Lambda CDM$ Cosmological Model. Data from missions like COBE, WMAP, and Planck confirmed the homogeneity and isotropy of the CMB within the order of one part in $10^5$, in line with the Cosmological Principle. Observations of large-scale structure, such as those from the Sloan Digital Sky Survey (SDSS), further corroborate the predictions made by FLRW cosmology, particularly regarding the distribution of galaxies.{\new While these experimental validations solidify the $\Lambda CDM$ model as the cornerstone of our current understanding of cosmology, recent data—such as from the Dark Energy Spectroscopic Instrument (DESI)—suggests that dark energy may not be best described solely by a cosmological constant. These observations open the possibility of alternative explanations, such as dynamic forms of dark energy or modifications to general relativity, which challenge the simplicity of the $\Lambda CDM$ framework while still adhering to the underlying assumptions of FLRW cosmology.}} 

Remarkably, however, when the story regarding the Friedman equations and the expansion of the universe is combined with the Dirac criterion for observables we run into an immediate and deeply problematic conflict. As just noted, the Friedmann equations describe the evolution of the scale factor $a$ and this appears to provide a clear description of the time evolution of the spatial geometric structure of the universe. However, the Friedmann equations are equivalent to those generated by a Hamiltonian constraint. So, \textit{if} the evolution equations generated by a constraint are interpreted as gauge transformations, \textit{then} we should understand the Friedmann equations not as dynamical equations, but as gauge equations.\footnote{We might, of course, simply reject Dirac's argument connecting gauge transformations to Hamiltonian constraints. In particular, his theorem that first class constraints generate gauge transformations does not apply to Hamiltonian constraints, see \citep{Barbour:2008} and \cite[\S 7.3]{Gryb:2023} for details. Moreover, rigorous formal analysis of these constraints indicate that there are distinct gauge generating and dynamics generating roles that can be explicitly disentangled in the case of theories with a single Hamiltonian constraint. See \cite[\S 13]{Gryb:2023} for details.} Quantities such as $a$ might appear to evolve over time in cosmology. However, they are not gauge-invariant, and so this evolution is not to be understood as physical. Remarkably, this conflict between the Friedman equations and the Dirac criterion for observables has received almost not detailed discussion in the physics or philosophy literature.  

The major exception is the discussion of \cite{Thiemann-k-essence}, who explicitly argues that `it is incorrect to interpret the FLRW equations as evolution equations of observable quantities' (p. 9) although he does `not doubt the validity the Einstein equations' he wants to `stress that their interpretation as physical evolution equations of observables is fundamentally wrong' (p. 9). Moreover on this view:  `All textbooks on classical GR incorrectly describe the Friedmann equations as physical evolution equations rather than what they really are, namely gauge transformation equations. The true evolution equations acquire possibly observable modifications to the gauge transformation equations whose magnitude depends on the physical clock that one uses to deparametrise the gauge transformation equations.' (p. 3). A simple approach to formalising this idea is to note that $a(t)$ does not Poisson commute with the Hamiltonian constraint of the theory. The time derivative of the scale factor is given by: $\dot{a}(t)=\{\mathcal{H},a\}\neq 0$, where $\mathcal{H}$ is the Hamiltonian constraint in canonical GR, and 
$\{,\}$ denotes the Poisson bracket. When the theory is understood in these terms it is indisputable that the scale factor is not a Dirac observable. This leads to an apparent contradiction with the physical observations of the universe's expansion. The natural response to this problem is to apply the complete and partial observable scheme to construct Dirac observables based upon the Friedmann system of equations. This is precisely what Thiemann suggests to do. His explicit proposal involves introducing a scalar field as a clock, capable of deparametrising the theory through the \cite{Brown1995} mechanism. This approach allows for the construction of a physical Hamiltonian, which generates the evolution of gauge-invariant Dirac observables. Since the dynamics of the chosen material system and its stress-energy tensor are taken into account, it is also to explicitly implement a deparametrisation in terms of a Real Reference Frames (\textbf{RRF}) as per the discussion of the last section. 

We give a sketch of the construction.
Let us introduce a spatially homogeneous scalar field $\phi$, which acts as a `phantom' field, which is not directly observable in modern cosmology but can have significant dynamical consequences leading to observable effects.\footnote{It is worth nothing here that there are two importantly different senses of phantom that coincide in Thiemann's usage. First, `phantom' in the sense of `missing physics' that is not directly observable in modern cosmology. Second, `phantom' in the more formal sense used by cosmologists as indicating a field with a first order kinetic term in the Lagrangian with a coefficient which has a sign opposite to the sign in the Klein–Gordon Lagrangian. See \cite[p. 4]{Thiemann-k-essence}.} The key innovation of this approach is to deparametrise the Hamiltonian constraint of GR, transforming it from a constraint equation into a physical Hamiltonian. The Hamiltonian constraint $\mathcal{H}$ is rewritten as: $\mathcal{H}=\pi+\mathtt{h}$, where
$\pi$ is the conjugate momentum to the scalar field and $\mathtt{h}$ is called the \textit{physical Hamiltonian} generating the temporal physical evolution of observables. Using this scalar field Thiemann constructs a framework in which the universe's time evolution is generated by the physical Hamiltonian rather than the Hamiltonian constraint. This reformulation allows for the deparametrisation of the theory, with $\mathtt{h}$ now acting as a physical Hamiltonian that generates time evolution for gauge-invariant observables. The crucial difference here is that $\mathtt{h}$ is not constrained to vanish, as is the case with the traditional Hamiltonian constraint in GR.

Once the theory is deparametrised, the time evolution of observables such as the scale factor \( a(\phi) \) can be computed using the physical Hamiltonian:
\begin{equation}
\frac{da(\phi)}{d\phi}|_{\phi=s} = \{ \mathtt{h}, a(\phi)\}|_{\phi=s} \neq 0
\end{equation}
for $s\in\mathbb{R}$. Of course,
\begin{equation}
\{ \mathcal{H}, a(\phi) \}|_{\phi=s} = 0.
\end{equation}
This evolution is now consistent with the Dirac criterion, as $a(\phi)$ is a Dirac observable that Poisson commutes with all constraints, unlike the original scale factor \( a(t) \), which did not Poisson commute with the Hamiltonian constraint.  Crucially, however, in this framework, the evolution of $a(\phi)$ is governed by a \textit{modified version} of the Friedmann equations, which includes additional terms due to the presence of the scalar field. In particular, the first Friedmann equations reads as:
\begin{equation}
\left( \frac{da/d\phi}{a(\phi)} \right)^2 = \left[ \cfrac{8\pi G}{3} \left[  \rho_m(\phi) + \rho_{\text{phantom}}(\phi) \right] +\cfrac{\Lambda}{3}  \right] \left( 1 + \frac{1}{x} \right),
\end{equation}
\noindent where $x = \frac{E^2}{\alpha^2 a(\phi)^6}$ is a deviation parameter, used to quantify how much the dynamics of the universe, governed by the modified Friedmann equation, differs from the standard cosmological model; $E$ is a constant of motion, representing the energy of the universe; $\alpha$ is a model parameter characterising the influence of the phantom field. We chose $k=0$ to adhere to Thiemann's formalism.

We thus arrive at an observationally distinct formulation of the theory which implements the Dirac observable prescription. The implication is then that we can either have the standard Friedmann equation and give up on our formalism for gauge-invariant observables or we can keep our formalism for observables gauge-invariant observables and modify the Friedmann equations. We cannot have both. Thiemann emphasises the gravity of this problem, stating that either the mathematical formalism of GR is inappropriate for cosmology, or we are missing some new physics. In the following section we seek to extricate ourselves from Thiemann's dilemma based upon the use of Einstein's famous clock hypothesis: that physical clocks measure proper time along their world-lines.

\section{Hubble Flow and the Clock Hypothesis }
\label{HFandCH}
Let us return to the derivation of the Friedman equations and seek an alternative physical interpretation of the $t$ in the equations. One crucial aspect of the model we have not yet explicitly considered is the idea of \textit{Hubble flow}. This \lq{}flow\rq{} describes the large-scale motion of matter, driven by the expansion of spacetime itself. More formally, Hubble flow is the component of recessional velocity of matter due to the expansion, separating it from peculiar velocities caused by local gravitational interactions. 

One way of understanding the derivation of the FLRW metric is via the adoption of the so-called synchronous reference frame \citep{Landau1987-fh}. In this frame there is a common, global cosmological time for all observers comoving with the Hubble flow. Crucially, this means that the synchronous frame is a \textit{geodetic reference frame}. Consequently, time-like trajectories orthogonal to space-like 3D hypersurfaces are geodesics of space-time and the four-velocity of each observer $U^\mu=(1,\Vec{0})$ automatically satisfy the geodesic equation. 

From this perspective the Friedmann equations are \textit{not} gauge transformation equations. Rather, they are evolution equations \textit{in} a specific gauge. We can see this as follows. Recall that in the Arnowitt-Deser-Misner (ADM) formalism \citep{Arnowitt1960} for canonical general relativity the metric is expressed in terms of the lapse function \(N(t)\), the shift vector \(N^i(t)\), and a spatial metric \(h_{ij}\).\footnote{From \cite{Geroch1970}'s Theorem follows that a globally hyperbolic spacetime can be foliated, that is decomposed into spatial slices parametrised by a global parameter \(t\). The ADM formalism can thus be applied to any globally hyperbolic spacetime.} The lapse function \(N\) formalises the temporal separation between two infinitesimally close hypersurfaces, measured in the normal direction to the first hypersurface. The shift vector \(N^i\) measures the displacement between the spatial coordinates \(x^i\) of a point \(P \in \Sigma_t\) and its orthogonal projection \(Q \in \Sigma_{t+dt}\). The connection between \(N\) and temporal diffeomorphisms and \(N^i\) and 3-diffeomorphisms emerges. It is specified that, in order to have a future-directed foliation, the lapse function \(N\) must be positive. In this formalism the general line element becomes:
\begin{equation}
ds^2 = -N^2 dt^2 + h_{ij} \left( dx^i + N^i dt \right) \left( dx^j + N^j dt \right).
\end{equation}
In the specific case of FLRW cosmology, the lapse function \(N(t) = 1\) and the shift vector \(N^i(t) = 0\). 

These choices defines the synchronous gauge, reflecting the absence of preferred locations and directions and simplifying the FLRW metric. Since in the synchronous gauge we can assume the coordinate time $t$ to coincide with the proper time $\tau$ measured by observers comoving with the Hubble flow, we will have that the $t$ in the equations will coincide with the proper time of the relevant bundle of geodesics following the Hubble flow. This means we can re-write the equations in terms of proper time $\tau$ simply by equating $t=\tau$.

It is worth noting that it is not necessary to show that $\{\mathcal{H},a(\tau)\}=0$, since we already `gauge-fixed' to the synchronous gauge. This means that $a(\tau)$ can be seen as a gauge-fixed observable and, as such, is gauge-invariant.
A gauge-fixed observable is defined in that particular gauge, and need not commute with the constraints because there are no more gauge transformations to refer to. In other words, the gauge freedom has already been eliminated and we no longer have any gauge constraints left, because we have chosen a specific reference frame (on this, see e.g. \citealp{WALLACE202438}). When can then clarify the claim of \cite[p.1914]{Dittrich2007} that complete observables and gauge-fixed observables \textit{are} the same, since a choice of reference frame can be seen as a gauge choice (a choice of section on a fibre bundle). Or, at least, it is always true that a gauge-fixed observable is a complete observable and only the reverse is less immediately obvious. For our current purpose, the key insight is recognising that the Friedmann equations, when considered in a synchronous gauge, function like dynamical evolution equations.\footnote{A more detailed discussion of the status of gauge-fixings, reference frames and gauge invariant observables in the context of cosmic proper time, within the fibre bundle formalism, is provided in Appendix Af. It is worth noting that conceptualising the Friedmann equations as gauge-fixed equations can be made consistent with the arguments of \cite[\S 13.2]{Gryb:2023} which imply a dynamical view of global Hamiltonian constraints in which the flow along a solution of the Friedmann equations would \textit{not} formally be a gauge symmetry. This is because one may view the symmetries of the theory as acting on entire histories and take their action on the boundary to be fixed and then recover the `gauge-fixed' perspective where the Friedmann equations in synchronous gauge describe evolution of the true degrees of freedom.}

To show this, consider that the Hubble flow should be understood to be the flow of a perfect fluid whose stress-energy tensor in the synchronous gauge is $T_{\mu\nu}=\text{diag}(\rho(\tau),-p(\tau))$, where $\rho(\tau)$ is the fluid's energy density, $p(\tau)$ its pressure. The Friedmann equations are then understood to describe how the expansion rate changes with the proper time of the observers comoving with the fluid. They can be written in terms of the Hubble parameter \(H(\tau) = \dot{a}(\tau)/a(\tau)\) as:
\begin{eqnarray}
H(\tau)^2 = \frac{8 \pi G}{3} \rho (\tau) - \frac{k c^2}{a(\tau)^2}  + \frac{\Lambda c^2}{3}\\
\dot{H}(\tau) + H(\tau)^2 = -\frac{4 \pi G}{3} \left( \rho(\tau) + \frac{3p(\tau)}{c^2} \right) + \frac{\Lambda c^2}{3}
\end{eqnarray}
where we have used the fact that $\frac{\ddot{a}}{a}=\dot{H} + H^2$ and assumed differentiation with respect to proper time. For a given matter model we can then write the density and pressure in terms of the scale factor allowing us to for example re-write the first equation as:
\begin{equation}
H^2(\tau) = H^2_0 \left[ \Omega_{m,0} \frac{1}{a(\tau)^3} + \Omega_{r,0} \frac{1}{a(\tau)^4} + \Omega_\Lambda + \Omega_k \frac{1}{a(\tau)^2} \right],
\end{equation}
where $H_0$ is the Hubble constant and we have introduced \textit{experimentally measurable} density parameters \textit{at the current time}; $\Omega_{0,R}$ for the radiation density, $\Omega_{0,M}$ for the matter (dark plus baryonic) density, $\Omega_{0,k}$ for the spatial curvature density, and $\Omega_{0,\Lambda}$ for the cosmological constant density.\footnote{The measurement of $H_0$ suffers from the so-called \textit{Hubble-tension}. For a review see \cite{Smeenk_2022}.}
%: a 5$\sigma$ discrepancy between the experimental values obtained from the method using Cepheids and Type Ia Supernovae as standard candles and the CMB measurements via PLANCK experiment. A 5$\sigma$ tension means that, statistically, the probability of the two measurements being compatible with each other is extremely low (less than one chance in a million).
The question is then whether and in what sense Hubble parameter $H(\tau)$ and the proper time along the Hubble flow can be understood to be \textit{measurable quantities}. Let us consider each in turn. 

First, the Hubble parameter. Experimentally, we cannot of course \textit{directly} measure the Hubble parameter with an \lq{}H-meter\rq{}.  However, we can surely measure it \textit{indirectly}, using astronomical observations that allow us to trace the expansion of the universe at different cosmic epochs. Let's just consider one of the various experimental possibilities. Type Ia supernovae are considered \textit{standard candles} in cosmology because they have a well-known absolute luminosity. By measuring the apparent luminosity of a supernova, we can estimate its distance (luminosity distance $d_L$), while the redshift $z$ is a direct indicator of how much the universe has expanded from the time $\tau$ of light emission of an object to the present day. The luminosity distance $d_L$ is related to the Hubble parameter $H(z)$ through the following relation:
\begin{equation}
d_L(z) = (1 + z) \int_0^z \frac{c \, dz'}{H(z')}\label{distance}
\end{equation}
By measuring the redshift $z$ through spectrometers and the distance $d_L$ through  large-field telescopes, such as the Hubble Space Telescope, we can infer $H(z)$. However, notice that this procedure requires \lq{}fitting\rq{} noisy discrete data, introducing mild model dependence. From $H(z)$ the value of the Hubble parameter $H(\tau)$ at different times $\tau$ can be derived.
In fact, using the definition of redshift in terms of the scale factor $a$, namely $a(\tau) = \frac{1}{1+z}$, and the fact that $H = \frac{\dot{a}}{a}$, we can derive the following expression:
\begin{equation}
H(z) = - \frac{\dot{z}}{1+z}.
\end{equation}
This equation highlights that, since it is impractical to directly measure $\dot{z}$, we must infer its behaviour using a specific cosmological model. The construction of $H(\tau)$ requires combining observations with a theoretical model that relates the cosmic time $\tau$ to the redshift $z$.\footnote{See Section \ref{sec5}, formula \eqref{time}, where a way to calculate $\tau(z)$ is provided.} Thus, any inference of $H$ from redshift measurements likely involves some degree of model dependence. The procedure to transition from $H(z)$ to $H(\tau)$ can be summarised as follows: I) Infer $H(z)$ using formula \eqref{distance}; II) Compute $\tau(z)$ according to the chosen theoretical model (see Section \ref{sec5}); III) Invert the relation $\tau(z)$ to obtain $z(\tau)$; IV) Combine $H(z)$ with $z(\tau)$ to construct $H(\tau)$. This procedure underscores the fact that $H$ and $\tau$ are not completely independent: knowledge of one requires knowledge of the other. Their (dynamical) interdependence supports the claim that $H(\tau)$ is a complete observable, in line with the remark outlined in Section \ref{COandRRF}, proposed by \cite{BamontiGomes2024}. For completeness, we note that model dependence is not unique to the measurement of $H(z)$. Many cosmological parameters rely on similar assumptions.

Second, and more subtle, is the question of whether we can measure the proper time along the Hubble flow. Again we evidently cannot measure \textit{directly} the proper time $\tau$ of our galaxy following the Hubble flow. If we say that we use \textit{whatever} periodic physical system as a clock `attached to the galaxy', it will not follow the Hubble flow. Actually, the concept of Hubble flow can be valid only at cosmological scales, so even for our Galaxy we should account for the effects of peculiar velocities. Completely eliminating peculiar motions from measurements of galaxy recession velocities is not possible, but it is possible to correct them in an approximate way. Therefore, strictly speaking, experimentally we cannot measure \textit{directly} with a clock the proper time of \textit{any} object following the Hubble flow. What we can do is appeal to the clock hypothesis. This amounts to the  \textit{hypothetical} assumption there is a clock that measures proper time along any given world-line. Proper time can be rigorously defined in the context of a relativistic spacetime $(M, g_{ab})$ as follows \cite[\S2.3]{malament2012topics}: let $\gamma : [s_1, s_2] \to M$ be a smooth future-directed timelike curve in the manifold $M$ with tangent $\xi^a$. Then the proper time associated with the curve relative to the metric $g_{ab}$ is given by:
\begin{equation}\label{propertime}
||\gamma|| = \int_{s_1}^{s_2} \left( g_{ab} \xi^a \xi^b \right)^{\frac{1}{2}} ds
\end{equation}
where $ds$ is the line element. Since the clock hypothesis applies also to the world-lines of observers following the Hubble flow, we seem to have solved the problem by stipulation. Since we can associate measurements with partial observables and partial observables with reference frames, we can wonder: what kind of Reference Frame the cosmic proper time is? And it is a \textit{bona fide} partial observable?
Given the clock hypothesis, we can stipulate a clock that measures cosmic time. Furthermore, cosmic time will always be dynamically coupled with the expansion rate of the Universe, parametrised by $H(\tau)$, since both quantities depend on the same FLRW metric. The cosmic time value, being a proper time, is a structural property of the gravitational field given by equation \eqref{propertime}. This is very similar to the sense in which the Hubble parameter measures the rate of expansion of the volume of the universe as a geometric quantity, which is derived from the gravitational field: $dV=\sqrt{-g}dx^\mu$. $\tau$ is a geodesic reference clock, since the four-velocity of the cosmological fluid is associated to the geodesic dynamics of a dust fluid, whose energy-momentum tensor $T_{ab}=\rho U_a U_b$ is source of the EFEs and give rise to the FLRW solution. Thus, $\tau$ is an \textbf{RRF} in \citep{Bamonti2023}'s classification. The nature of the \textbf{RRF} clock comes from the fact that $\tau$ is the proper time of the cosmological fluid, whose back-reaction on gravity is taken into account and gives rise to the FLRW metric which in turn determines the proper time $\tau$ (this is the essence of the non-linear feedback of EFEs).
This \textbf{RRF} clock provides the privileged representation in which the cosmic microwave background radiation is represented as perfectly homogeneous and isotropic, in absence of small inhomogeneities of the primordial universe. We thus have that cosmic proper time: i) is an \textbf{RRF} since it involved back-reaction; ii) is a \textit{bona fide} partial observable according to \cite{BamontiGomes2024}; and iii) corresponds to a measurable quantity by the clock hypothesis. Is this enough for us to conclude that $H(\tau)$ is a complete observable and thus have solved Thiemann's dilemma? Almost.

As stated above, the problem is that \textit{on a practical level}, it is not possible to have an experimentally accessible clock (i.e. with which we can exchange signals), that follows the Hubble flow.  In general, distant galaxies are considered to follow the Hubble Flow, as for very distant galaxies, the contribution of their peculiar motions is negligible compared to the recession velocity due to the expansion of the universe.
Thus, let's consider a galaxy in our past light cone as a satellite sending radiation towards us. In this way, we would use it as a kind of \cite{RovelliGPS}'s ‘GPS clock\rq{} that would allow us to define local quantities, such as the Hubble parameter, in its proper time.\footnote{Note also that clusters of matter represent inhomogeneities that are assumed to evolve following the underlying FLRW background structure. So, their evolution does not influence the global FLRW evolution. \lq{}\lq{}More precisely, it is assumed that effects from the small scale inhomogeneities onto the largest scales can be neglected, i.e. there is no substantial backreaction\rq{}\rq{} \cite{Schander2021}. Thus, galaxies, clusters and other agglomerates of matter are treated as \textit{test particles}.}
Again, however, the cosmic time would be the time measured by some stipulated and not further defined clock ‘attached’ to that galaxy-satellite and broadcasting the measured value to the experimenter, via light signals. Apart from the experimental problem to construct a valuable experimental setting, and the need to take into account the galaxy's peculiar velocities, there remains the problem of defining an origin of such time measured by the clock-galaxy.
One solution would be to conveniently place the origin as the ‘zero time’ of the formation event of such a galaxy. Or also, analogous to the construction of GPS coordinates in \cite{RovelliGPS}, as the time of the galaxy's encounter with another galaxy.
In any case, the proper time of the galaxy will never be a global time. As Rovelli states: `Our Galaxy and Andromeda are heading towards a collision: when they will meet, the times elapsed from the Big Bang will be different in the two galaxies. None of the two will have any claim of being more of a ``true'' time than the other.' \cite[p.18]{RovelliPrinceton}. It is practically impossible to have a clock measuring the proper time parametrising the Hubble flow.\footnote{In a similar vein, \cite{brown:2016}, note that \lq{}\lq{}For any given clock, no matter how ideal its performance when inertial, there will in principle be an acceleration-producing external force, or even tidal effects inside the clock, such that the clock ``breaks'', in the sense of violating the clock hypothesis. Might it not be more appropriate to call it the \textit{clock condition}?\rq{}\rq{}} In this context, one might view the clock hypothesis as constituting the \textit{definition} of clocks as \textit{objects that measure proper time}. The important point is that \textit{if} the clock hypothesis holds, then we are able to treat the $\tau$ in the Friedmann equations as a measurable quantity just like the Hubble parameter. However, in the context of cosmology and the Hubble flow, there is a tension between applying the clock hypothesis and the idea of a clock as a real, experimentally accessible physical system. There do not exist real physical systems that approximate clocks which measure proper time along the Hubble flow. In the following section we return to the ideas of partial and complete observable and \textbf{RRFs} to better understand both this challenge and the comparative merits of the approach of Thiemann described in the previous section. 

\section{Finding Cosmic Time}\label{sec5}

Let us recap. One the one hand, the widely used and accepted  criterion for an observable in a theory with temporal diffeomorphism symmetry is that such observables should be Dirac observables and therefore have (weakly) vanishing Poisson bracket with all first class constraints. On the other hand, the widely used and empirically established Friedmann equations describing the dynamics of the scale factor can be understood to correspond to those generated by a first class constraint in a theory with temporal diffeomorphism symmetry. It seems like we must either give up on the Dirac criterion for observables or modify our understanding cosmological dynamics. We have considered two alternative responses to this dilemma as follows. First, Thiemann argues that we should adopt the second option and demonstrates how we might explicitly reconstruct Friedmann cosmology as a de-parametrised theory based upon a phantom matter field acting as a physical clock that measures cosmic time. A deviation parameter then quantifies how much the dynamics of the universe, governed by the modified Friedmann equation, differs from the standard model of cosmology. Second, we have constructed an alternative approach that re-interprets Friedmann equations as evolution equations parametrised by proper time, rather than coordinate time. On this approach we understand the equations as describing the correlation between two independently measurable `partial observables' given by the Hubble parameter and proper time along the Hubble flow. The Hubble parameter is a measurable quantity within modern cosmology. Furthermore, following the clock hypothesis, we have that since clocks measure proper time along world-lines, a clock following the Hubble flow will necessarily  measure cosmic time. The problem, however, is that the proper time along the Hubble flow does not correspond to a physical quantity associated with a measuring procedure by a clock leading to a number, and so it seems we are no longer implementing the partial and complete observables procedure in the spirit of \cite{Rovelli_2002}. The crucial issue is to define an experimental measure of cosmic time consistent with the clock hypothesis. 

In this context, it is worth noting again that Thiemann holds a different understanding of the partial and complete observables approach to Rovelli. In particular, for Thiemann partial `observables' are not observables at all and so there is no sense in which they need to be associate to a measuring procedure. We can thus understand the Thiemann approach to complete observables and cosmic time as built upon abandoning not one by two conventionally accepted aspects of the formalism, viz. the clock hypothesis and the distinction between partial and complete observability. An \textit{observable simpliciter} is defined as a Dirac Observable and there is no requirement that such observables are built out of independently measurable functions, which however are not (Dirac) observables. On this way of thinking, $H(\tau)$ will \textit{not} be an \textit{observable simpliciter}, and should not be expected to commute with the Hamiltonian constraint. 
In fact, since the clock hypothesis is abandoned, $\tau$ is not observable, and for that $H(\tau)$ is not a (Dirac) observable either. Plausibly, it is precisely the abandonment of the clock hypothesis which led Thiemann to use a phantom scalar clock as the physical, observable clock of the theory. In any case, the crucial point is that adopting this perspective does not amount simply to an alternative interpretation of the theory. Rather, it is to reformulate the classical theory of cosmology such in a way that modifies the empirical consequences. What modifications are made will depend upon the clock choice. However, there is no choice that corresponds to a strict preservation of the Friedmann equations:  `whatever matter is used for deparametrisation, there will be corrections [...] This should have observable consequences!'  \cite[p.9]{Thiemann-k-essence}. Non-standard empirical consequences are of course a virtue in a physical theory. Furthermore it is worth noting that Thiemann's approach in the paper in question is connected to a specific research programme in terms of Phantom k-essence cosmology \citep{aguirregabiria:2004} and thus the modifications in question could be independently motivated and in principle tested via the relevant modified matter or gravity theory. There are thus good methodological reasons to pursue more radical approach to identifying cosmic time.

Can we plot a plausible path towards a more conservative approach that preserves both the distinct notions of partial and complete observable and the clock hypothesis? Recall once more that on the original Rovelli's definition a partial observable is a physical quantity associated with \textit{measuring procedure} leading to a number. This definition fits well with the way the Hubble parameter features in our cosmological observational practice, even if its measurement is \textit{indirect}. The problem was that proper time along the Hubble flow is not associated with a \textit{direct} measuring procedure in any straightforward sense that involves a clock. Notwithstanding the clock hypothesis, there seems to be a tension between our intuitive notion of measurement of time (by a clock) and the role played by cosmic proper time in our scientific theories. It is worth considering at this juncture, however, that the intuitive notion of measurement has itself been disputed in the context of a practice orientated account of scientific measurement. This has lead to a reorientation of the philosophy of scientific measurement towards a \textit{model-based account of measurement}. Furthermore, perhaps the most detailed and powerful example of the conceptual heavy lifting that a model-based account of measurement can do is in context of the measurement of time. 

Let us briefly consider key ideas from the path-breaking work of \cite{tal:2016} on the measurement of time via atomic clocks cf. \cite{thebault:2021b}. Following this account we recognise that the time `measured' by atomic clocks does not correspond to a simple procedure of reading a number from a device. Rather, `Coordinated Universal Time' (UTC) is based upon a \textit{standardisation procedure} involving multiple atomic clocks distributed throughout the globe and systematic modelling at various stages. Caesium plays a particularly important role in modern time-keeping since it is transitions of an \textit{idealised} caesium atom that are the basis for the definition of the second.
However, as emphasised by Tal, this does not mean that one can simply read seconds from real caesium atoms. The caesium atom that defines the second is an idealised construct, at rest at zero degrees Kelvin and with no coupling to any external fields.  Actual atomic clocks are built to approximately realise the ideal caesium clock, with \textit{known} sources of difference minimised and modelled.  However, the `primary' standards caesium clocks typically only operates for a few weeks at a time in order to calibrate `secondary standards'. The secondary standards are a different class of atomic clocks that are less accurate but can be run continuously for a number of years. The secondary standard clocks also must be modelled. In particular, the `readings' of the clocks are subject to quantitive adjustments relating to the known sources of difference between their ideal physical operation and their actual physical realisation. This allows the time that they read to be a close approximation to that read by their idealised counterpart. The crucial point is that UTC is not `read' by either primary or secondary standards. Rather it is a product of a further abstraction based upon the readings of the different participant clocks throughout the world. Furthermore, not only are different clocks weighted differently in UTC, since some clocks are more noisy, but since the clocks are at different physical locations on the earth, one must also take account of their differing proper times, as determined by the relevant differences in gravitational field and four-acceleration \citep[p. 302]{tal:2016}. One also requires a procedure for synchronisation which introduces an element of conventionality. However, despite including conventional elements, the synchronisation of the clock network towards UTC is partially anchored in underlying regularities in nature which are required for the explanation of successful stabilisation of a synchronisation standards in \lq{}metrological practice\rq{} \cite[\S 3.1]{tal:2016}.\footnote{For specific technical details regarding the relativity of synchronisation in the context of UTC and the GPS system see the discussion of \cite{ashby:2003}.}
  
The general implication for a model-based account of measurement for the definition of a partial observable is thus to substantially liberalise the sense of `procedure' in Rovelli's definition. In particular, on a model-based account of measurement, the procedure involved may include not just indirect measurement but various tiers of modelling, calibration, and aggregation. Furthermore, and more importantly for our discussion, comparison between the status of UTC and cosmic time throws into relief our failure in the latter case to find an actual physical system that plays the role of the clock. \textit{There are no actual physical systems that measure UTC either.}
Consider, then, that we can `measure' cosmic time via various cosmological phenomena. For example, we can use the analysis of the power spectrum of temperature fluctuations of the CMB (TT-power spectrum) measured by the PLANCK satellite. In particular, the CMB power spectrum represents the power distribution of the temperature anisotropies as a function of angular scale. The different angular scales correspond to the scales of the baryonic acoustic oscillations (BAOs) which are pressure waves generated by interactions between radiation (photons) and matter (baryons) in the primordial plasma before decoupling era. Anisotropies reflect primordial density differences, which led to the formation of galaxies and other structures \cite[ch.9]{kolb1994early}. The power spectrum contains peaks and valleys at different scales and their position and amplitude are sensitive to cosmological parameters. For example, a higher density of matter leads to higher peaks that are closer together, and a flat universe tends to have a different distribution of peaks than a curved universe. The  Hubble constant $H_0$ also affects the scale of the fluctuations and the position of the peaks.
%\footnote{Formally, the TT-power spectrum is the expansion in Legendre polynomials $\mathfrak{P}$ of the variance of temperature fluctuations: $$\langle \cfrac{\Delta T}{T}(\gamma_1)\cfrac{\Delta T}{T}(\gamma_2)\rangle=1/{2\pi}\sum_l{l(l+1)C_l \mathfrak{P}_l(cos\theta)},$$ where $\gamma$ represents the direction in which temperature fluctuations $\Delta T$ are measured and $C_l$ is the coefficient of the $l$-th multipole, representing the amplitude of temperature fluctuations at a certain angular scale $\theta \propto 1/l$. There is a privileged angular scale for these oscillations, which corresponds to the size of the sound horizon at the time of recombination. This angular scale corresponds approximately to the multipole $l = 200$ with a dependence on cosmological parameters.} See Figure 1 in  \cite{bamonti:2024}.

By reconstructing the power spectrum it is possible to determine cosmological time. In particular, cosmological time can be determined by integrating the equations of the expansion of the universe:
\begin{equation}
\tau(z) = \int_{z}^{\infty} \frac{dz'}{(1+z') H(z')}, \label{time}
\end{equation}
where $H(z)$ is obtained from the cosmological parameters $\Omega_I,H_0$ which are determined by the experimental power spectrum.
To be precise, $\tau(z)$ above does not correspond to the proper time of a real observer comoving with the Hubble Flow, since in observational practice, we know the Universe is not \textit{perfectly} homogeneous and isotropic: not even on cosmologically large scales. The FLRW model is an idealisation to describe the average behaviour of the Universe.
The Hubble parameter \(H(z)\) in the formula depends on parameters such as \(\Omega_{\text{CDM}}\) (dark matter density), which are in part determined using perturbative methods, which take into account small inhomogeneities.
Therefore, in measuring the parameter \(H(z)\), contributions of inhomogeneity (such as galaxy clusters, voids) and anisotropies are taken into account.
Nonetheless, the value of \(H(z)\) used in the theoretical calculation of $\tau(z)$ refers to the \lq{}average\rq{} behaviour of the Universe, described by a homogeneous and isotropic model.
We mean that the formula \(\tau(z)\) calculating cosmic time is based on a model that idealises the Universe as perfectly homogeneous and isotropic, using an \lq{}average value\rq{} of \(H(z)\).
This means that the cosmic time $\tau(z)$ is a global average, which does not reflect local fluctuations or deviations from isotropy, but rather provides an approximated estimate.
One could say that, experimentally, the contributions from inhomogeneities and anisotropies of the actual Universe are \lq{}\textit{averaged out to zero}\rq{}, in the sense that these contributions are present in the data but, when calculating the Hubble parameter \(H(z)\), local fluctuations are essentially \lq{}smoothed out\rq{} and do not significantly influence the overall result.
Consequently, experimental measurements \textit{approximate} the theoretical ideal cosmic time, which refers to an ideally homogeneous and isotropic Universe, analogously to what happens with primary and secondary clocks for measuring UTC.
It is the case that $\tau(z)$ that is used to make predictions in cosmology, and it is a partial observable, pace \cite{Thiemann-k-essence}. We thus have that in a more liberalised sense of `measurement procedure', it \textit{is} the case that we can treat proper time along the Hubble flow as a partial observable in the context of cosmology. This has not involved simply stipulating proper time as partial observable via the clock hypothesis. However, it also has not involved the abandonment of the clock hypothesis altogether. Rather, the clock hypothesis reemerges within cosmology as something like a `coordinative definition' in sense of \cite{reichenbach:1928}. That is, the clock hypothesis allows for the \textit{coordination} of a concept (an ideal clock measuring the Hubble flow temporal parametrisation) with an empirical phenomenon (cosmic time). This broadly logical empiricist understanding accords with other discussions of the clock hypothesis in the recent philosophical literature \citep{adlam:2022} and makes sense of the physically non-trivial but partially definitional role, cf. \citep{Fletcher}. We recover the Friedmann equations and cosmic time whilst keeping both the clock hypothesis and the partial/complete observable distinction, albeit each in modified form. The conservative option for finding cosmic time is thus a live possibility.

\section{Summary and Outlook}

The complete and partial observables programme is a response to the problem of time that seeks to preserve the Dirac criterion for observables. Recent work on reference frames allows us to disambiguate the definition of a partial observable and better understand what it means for a physical variable to play the role of a clock in the context of a complete observable. Thiemann's approach to the interpretation of the Friedmann equations as dynamical equations leads to their de-parametrisation via a phantom field through the \cite{Brown1995} mechanism. An alternative approach is based on the idea of comic time as a proper time parameter along the Hubble flow. In this context, we can re-consider the observability and measurability of the Hubble parameter and its status as a complete observable whose dynamics is described by the unmodified Friedman equations.  The problem is then that there is no physical system which can be understood to \textit{even approximately} measure proper time along the Hubble flow. This leads to the introduction of a more liberalised sense of `measurement procedure' in the context of cosmic time drawing upon the model-based account of  measurement. Provided one is willing to reinterpret our criteria of observability in light of the clock hypothesis and the model-based account of measurement, one can preserve the Friedmann equations and find time in cosmology. 

The problem of time is often presented as arcane question in the foundations of quantum gravity. If the loss of time within our physical theories were only an issue at the Planck scale then it would hardly seem an urgent one. We hope to have given good reasons to dispel such complacency. In particular, we have show how aspects of the problem occur even in the familiar context of classical cosmology and even at scales upon which we already have a wealth of observable evidence, such as those relevant to the age of the universe. The way in which we approach the problem can have empirical physical consequences. Moreover, on our account, the cosmic problem of time is closely connected to both important methodological questions relating to idealisation and the nature of measurement and to familiar foundational topics such as the interpretation of the clock hypothesis and the nature of gauge degrees of freedom. As such, we suggest the search for cosmic time warrants more attention from both philosophers of science and scientists alike.  

\section{Acknowledgements}
We are extremely grateful to David Sloan, two anonymous referees, and an editor for helpful comments on draft manuscript(s). KT would also like to thank Sean Gryb for helpful discussion. 

\bibliographystyle{chicago}
\bibliography{BIB.bib}

\begin{thebibliography}{}

\bibitem[\protect\citeauthoryear{Adlam, Linnemann, and Read}{Adlam
  et~al.}{2022}]{adlam:2022}
Adlam, E., N.~Linnemann, and J.~Read (2022).
\newblock Constructive axiomatics in spacetime physics part ii: Constructive
  axiomatics in context.
\newblock {\em arXiv preprint arXiv:2211.05672\/}.

\bibitem[\protect\citeauthoryear{Aguirregabiria, Chimento, and
  Lazkoz}{Aguirregabiria et~al.}{2004}]{aguirregabiria:2004}
Aguirregabiria, J.~M., L.~P. Chimento, and R.~Lazkoz (2004).
\newblock Phantom k-essence cosmologies.
\newblock {\em Physical Review D\/}~{\em 70\/}(2), 023509.

\bibitem[\protect\citeauthoryear{Anderson}{Anderson}{2017}]{anderson:2017}
Anderson, E. (2017).
\newblock {\em The Problem of Time: Quantum Mechanics Versus General
  Relativity}, Volume 190.
\newblock Springer.

\bibitem[\protect\citeauthoryear{Anderson et~al.}{Anderson
  et~al.}{2014}]{anderson:2014}
Anderson, E. et~al. (2014).
\newblock Beables/observables in classical and quantum gravity.
\newblock {\em SIGMA\/}~{\em 10\/}(092), 092.

\bibitem[\protect\citeauthoryear{Anderson}{Anderson}{1967}]{Anderson1967-en}
Anderson, J.~L. (1967, May).
\newblock {\em Principles of relativity physics}.
\newblock San Diego, CA: Academic Press.

\bibitem[\protect\citeauthoryear{Andersson, Galloway, and Howard}{Andersson
  et~al.}{1998}]{andersson:1998}
Andersson, L., G.~J. Galloway, and R.~Howard (1998).
\newblock The cosmological time function.
\newblock {\em Classical and quantum gravity\/}~{\em 15\/}(2), 309.

\bibitem[\protect\citeauthoryear{Arnowitt et~al.}{Arnowitt
  et~al.}{1960}]{Arnowitt1960}
Arnowitt, R. et~al. (1960, March).
\newblock Canonical variables for general relativity.
\newblock {\em Physical Review\/}~{\em 117\/}(6), 1595--1602.

\bibitem[\protect\citeauthoryear{Arnowitt, Deser, and Misner}{Arnowitt
  et~al.}{1959}]{ADMII}
Arnowitt, R., S.~Deser, and C.~W. Misner (1959).
\newblock Dynamical structure and definition of energy in general relativity.
\newblock {\em Physical Review\/}~{\em 116\/}(5), 1322.

\bibitem[\protect\citeauthoryear{Arnowitt, Deser, and Misner}{Arnowitt
  et~al.}{1962}]{ADMReview}
Arnowitt, R., S.~Deser, and C.~W. Misner (1962).
\newblock {The Dynamics of General Relativity}.
\newblock In L.~Witten (Ed.), {\em Gravitation: An Introduction to Current
  Research}, Chapter~7, pp.\  227--265. New York: John Wiley \& Sons Inc.

\bibitem[\protect\citeauthoryear{Ashby}{Ashby}{2003}]{ashby:2003}
Ashby, N. (2003).
\newblock Relativity in the global positioning system.
\newblock {\em Living Reviews in relativity\/}~{\em 6}, 1--42.

\bibitem[\protect\citeauthoryear{Bamonti}{Bamonti}{2023}]{Bamonti2023}
Bamonti, N. (2023).
\newblock What is a reference frame in general relativity?
\newblock {\em https://arxiv.org/abs/2307.09338\/}.

\bibitem[\protect\citeauthoryear{Bamonti and Gomes}{Bamonti and
  Gomes}{2024}]{BamontiGomes2024}
Bamonti, N. and H.~Gomes (2024).
\newblock What reference frames teach us. part i: About symmetry principles and
  observability.
\newblock {\em https://arxiv.org/abs/2410.12892\/}.

\bibitem[\protect\citeauthoryear{Bamonti and Th{\'e}bault}{Bamonti and
  Th{\'e}bault}{2024}]{bamonti:2024}
Bamonti, N. and K.~P. Th{\'e}bault (2024).
\newblock In search of cosmic time: Complete observables and the clock
  hypothesis.
\newblock {\em arXiv preprint arXiv:2411.00541\/}.

\bibitem[\protect\citeauthoryear{Barbour}{Barbour}{2000}]{barbour2000end}
Barbour, J. (2000).
\newblock {\em The End of Time: The Next Revolution in Physics}.
\newblock Oxford, UK: Oxford University Press.

\bibitem[\protect\citeauthoryear{{Barbour} and {Foster}}{{Barbour} and
  {Foster}}{2008}]{Barbour:2008}
{Barbour}, J. and B.~Z. {Foster} (2008, August).
\newblock {Constraints and gauge transformations: Dirac's theorem is not always
  valid}.
\newblock {\em ArXiv e-prints\/}.

\bibitem[\protect\citeauthoryear{Barbour}{Barbour}{2001}]{barbour2001discovery}
Barbour, J.~B. (2001).
\newblock {\em The Discovery of Dynamics: A Study from a Machian Point of View
  of the Discovery and the Structure of Dynamical Theories}.
\newblock Oxford, UK: Oxford University Press.

\bibitem[\protect\citeauthoryear{Belot}{Belot}{2007}]{Belot:2007}
Belot, G. (2007, Jan).
\newblock The representation of time and change in mechanics.
\newblock In J.~Butterfield and J.~Earman (Eds.), {\em Handbook of Philosophy
  of Physics}, Chapter~2. Elsevier.

\bibitem[\protect\citeauthoryear{Belot and Earman}{Belot and
  Earman}{2001}]{belot_earman_2001}
Belot, G. and J.~Earman (2001).
\newblock {\em Pre-Socratic quantum gravity}, pp.\  213--255.
\newblock Cambridge University Press.

\bibitem[\protect\citeauthoryear{Bergmann and Komar}{Bergmann and
  Komar}{1962}]{Bergmann:1962}
Bergmann, P. and A.~Komar (1962).
\newblock Status report on the quantization of the gravitational field.
\newblock In {\em Recent Developments in General Relativity}, pp.\  31--46.
  Pergamon Press.

\bibitem[\protect\citeauthoryear{Bergmann}{Bergmann}{1956}]{bergmann:1956}
Bergmann, P.~G. (1956).
\newblock Introduction of true observables into the quantum field equations.
\newblock {\em Il Nuovo Cimento (1955-1965)\/}~{\em 3\/}(6), 1177--1185.

\bibitem[\protect\citeauthoryear{Bergmann}{Bergmann}{1961a}]{Bergmann61b}
Bergmann, P.~G. (1961a, Oct).
\newblock "gauge-invariant" variables in general relativity.
\newblock {\em Phys. Rev.\/}~{\em 124}, 274--278.

\bibitem[\protect\citeauthoryear{Bergmann}{Bergmann}{1961b}]{Bergmann1961}
Bergmann, P.~G. (1961b, October).
\newblock Observables in general relativity.
\newblock {\em Reviews of Modern Physics\/}~{\em 33\/}(4), 510--514.

\bibitem[\protect\citeauthoryear{Bergmann and Komar}{Bergmann and
  Komar}{1960}]{Bergmann:1960}
Bergmann, P.~G. and A.~B. Komar (1960).
\newblock Poisson brackets between locally defined observables in general
  relativity.
\newblock {\em Physical Review Letters\/}~{\em 4\/}(8), 432.

\bibitem[\protect\citeauthoryear{Bojowald, H{\"o}hn, and Tsobanjan}{Bojowald
  et~al.}{2011a}]{Bojowald:2011}
Bojowald, M., P.~A. H{\"o}hn, and A.~Tsobanjan (2011a).
\newblock An effective approach to the problem of time.
\newblock {\em Classical and Quantum Gravity\/}~{\em 28\/}(3), 035006.

\bibitem[\protect\citeauthoryear{Bojowald, H{\"o}hn, and Tsobanjan}{Bojowald
  et~al.}{2011b}]{bojowaldhohn:2011b}
Bojowald, M., P.~A. H{\"o}hn, and A.~Tsobanjan (2011b).
\newblock Effective approach to the problem of time: general features and
  examples.
\newblock {\em Physical Review D\/}~{\em 83\/}(12), 125023.

\bibitem[\protect\citeauthoryear{Bridgman}{Bridgman}{1927}]{bridgman:1927}
Bridgman, P. (1927).
\newblock The logic of modern physics.
\newblock {\em Beaufort Brooks\/}.

\bibitem[\protect\citeauthoryear{Brown and Read}{Brown and
  Read}{2016}]{brown:2016}
Brown, H.~R. and J.~Read (2016).
\newblock Clarifying possible misconceptions in the foundations of general
  relativity.
\newblock {\em American Journal of Physics\/}~{\em 84\/}(5), 327--334.

\bibitem[\protect\citeauthoryear{Brown and Kucha{\v{r}}}{Brown and
  Kucha{\v{r}}}{1995}]{Brown1995}
Brown, J.~D. and K.~V. Kucha{\v{r}} (1995, May).
\newblock Dust as a standard of space and time in canonical quantum gravity.
\newblock {\em Physical Review D\/}~{\em 51\/}(10), 5600--5629.

\bibitem[\protect\citeauthoryear{Callender and McCoy}{Callender and
  McCoy}{2021}]{callender:2021}
Callender, C. and C.~D. McCoy (2021).
\newblock Time in cosmology.
\newblock In {\em The Routledge Companion to Philosophy of Physics}, pp.\
  707--718. Routledge.

\bibitem[\protect\citeauthoryear{Casadio, Chataignier, Kamenshchik, Pedro,
  Tronconi, and Venturi}{Casadio et~al.}{2024}]{casadio:2024}
Casadio, R., L.~Chataignier, A.~Y. Kamenshchik, F.~G. Pedro, A.~Tronconi, and
  G.~Venturi (2024).
\newblock Relaxation of first-class constraints and the quantization of gauge
  theories: from" matter without matter" to the reappearance of time in quantum
  gravity.
\newblock {\em arXiv preprint arXiv:2402.12437\/}.

\bibitem[\protect\citeauthoryear{Chang}{Chang}{2021}]{sep-operationalism}
Chang, H. (2021).
\newblock {Operationalism}.
\newblock In E.~N. Zalta (Ed.), {\em The {Stanford} Encyclopedia of
  Philosophy\/} ({F}all 2021 ed.). Metaphysics Research Lab, Stanford
  University.

\bibitem[\protect\citeauthoryear{Colg{\'a}in, Dainotti, Capozziello,
  Pourojaghi, Sheikh-Jabbari, and Stojkovic}{Colg{\'a}in
  et~al.}{2024}]{colgain:2024}
Colg{\'a}in, E.~{\'O}., M.~G. Dainotti, S.~Capozziello, S.~Pourojaghi,
  M.~Sheikh-Jabbari, and D.~Stojkovic (2024).
\newblock Does desi 2024 confirm $lambda $ cdm?
\newblock {\em arXiv preprint arXiv:2404.08633\/}.

\bibitem[\protect\citeauthoryear{Dirac}{Dirac}{1950}]{dirac:1950}
Dirac, P. A.~M. (1950).
\newblock Generalized hamiltonian dynamics.
\newblock {\em Canadian journal of mathematics\/}~{\em 2}, 129--148.

\bibitem[\protect\citeauthoryear{Dirac}{Dirac}{1958a}]{Dirac:1958a}
Dirac, P. A.~M. (1958a).
\newblock Generalized hamiltonian dynamics.
\newblock {\em Proceedings of the Royal Society of London. Series A,
  Mathematical and Physical Sciences\/}~{\em 246}, 333--3343.

\bibitem[\protect\citeauthoryear{Dirac}{Dirac}{1958b}]{Dirac:1958b}
Dirac, P. A.~M. (1958b).
\newblock The theory of gravitation in hamiltonian form.
\newblock {\em Proceedings of the Royal Society of London. Series A,
  Mathematical and Physical Sciences\/}~{\em 246}, 333--343.

\bibitem[\protect\citeauthoryear{Dirac}{Dirac}{1964}]{Dirac:1964}
Dirac, P. A.~M. (1964).
\newblock {\em Lectures on quantum mechanics}.
\newblock Dover Publications.

\bibitem[\protect\citeauthoryear{DiSalle}{DiSalle}{2020}]{sep-spacetime-iframes}
DiSalle, R. (2020).
\newblock {Space and Time: Inertial Frames}.
\newblock In E.~N. Zalta (Ed.), {\em The {Stanford} Encyclopedia of
  Philosophy\/} ({W}inter 2020 ed.). Metaphysics Research Lab, Stanford
  University.

\bibitem[\protect\citeauthoryear{Dittrich}{Dittrich}{2006}]{Dittrich2006}
Dittrich, B. (2006, October).
\newblock Partial and complete observables for canonical general relativity.
\newblock {\em Classical and Quantum Gravity\/}~{\em 23\/}(22), 6155--6184.

\bibitem[\protect\citeauthoryear{Dittrich}{Dittrich}{2007}]{Dittrich2007}
Dittrich, B. (2007, August).
\newblock Partial and complete observables for hamiltonian constrained systems.
\newblock {\em General Relativity and Gravitation\/}~{\em 39\/}(11),
  1891--1927.

\bibitem[\protect\citeauthoryear{Dittrich, H{\"o}hn, Koslowski, and
  Nelson}{Dittrich et~al.}{2017}]{dittrich:2017}
Dittrich, B., P.~A. H{\"o}hn, T.~A. Koslowski, and M.~I. Nelson (2017).
\newblock Can chaos be observed in quantum gravity?
\newblock {\em Physics Letters B\/}~{\em 769}, 554--560.

\bibitem[\protect\citeauthoryear{Earman}{Earman}{1974}]{Earman1974}
Earman, J. (1974, June).
\newblock Covariance, invariance, and the equivalence of frames.
\newblock {\em Foundations of Physics\/}~{\em 4\/}(2), 267--289.

\bibitem[\protect\citeauthoryear{Earman and Friedman}{Earman and
  Friedman}{1973}]{Earman1973}
Earman, J. and M.~Friedman (1973, September).
\newblock The meaning and status of newton's law of inertia and the nature of
  gravitational forces.
\newblock {\em Philosophy of Science\/}~{\em 40\/}(3), 329--359.

\bibitem[\protect\citeauthoryear{Fankhauser and D{\"u}rr}{Fankhauser and
  D{\"u}rr}{2021}]{fankhauser:2021}
Fankhauser, J. and P.~M. D{\"u}rr (2021).
\newblock How (not) to understand weak measurements of velocities.
\newblock {\em Studies in History and Philosophy of Science Part A\/}~{\em 85},
  16--29.

\bibitem[\protect\citeauthoryear{Fletcher}{Fletcher}{2025}]{Fletcher}
Fletcher, S.~C. (2025).
\newblock {\em Foundations of General Relativity}.
\newblock Cambridge Elements in the Philosophy of Physics. Cambridge University
  Press (Forthcoming).

\bibitem[\protect\citeauthoryear{Friedman}{Friedman}{1922}]{Friedman1922}
Friedman, A. (1922, December).
\newblock Uber die krummung des raumes.
\newblock {\em Zeitschrift fur Physik\/}~{\em 10\/}(1), 377--386.

\bibitem[\protect\citeauthoryear{Gambini and Porto}{Gambini and
  Porto}{2001}]{gambini:2001}
Gambini, R. and R.~A. Porto (2001).
\newblock Relational time in generally covariant quantum systems: four models.
\newblock {\em Physical Review D\/}~{\em 63\/}(10), 105014.

\bibitem[\protect\citeauthoryear{Gambini, Porto, Pullin, and Torterolo}{Gambini
  et~al.}{2009}]{gambini:2009}
Gambini, R., R.~A. Porto, J.~Pullin, and S.~Torterolo (2009).
\newblock Conditional probabilities with dirac observables and the problem of
  time in quantum gravity.
\newblock {\em Physical Review D\/}~{\em 79\/}(4), 041501.

\bibitem[\protect\citeauthoryear{Geroch}{Geroch}{1970}]{Geroch1970}
Geroch, R. (1970, February).
\newblock Domain of dependence.
\newblock {\em Journal of Mathematical Physics\/}~{\em 11\/}(2), 437--449.

\bibitem[\protect\citeauthoryear{Giacomini}{Giacomini}{2021}]{Giacomini2021}
Giacomini, F. (2021, July).
\newblock Spacetime quantum reference frames and superpositions of proper
  times.
\newblock {\em Quantum\/}~{\em 5}, 508.

\bibitem[\protect\citeauthoryear{Gomes, Gryb, and Koslowski}{Gomes
  et~al.}{2011}]{gomes:2011}
Gomes, H., S.~Gryb, and T.~Koslowski (2011).
\newblock Einstein gravity as a 3d conformally invariant theory.
\newblock {\em Classical and Quantum Gravity\/}~{\em 28\/}(4), 045005.

\bibitem[\protect\citeauthoryear{Gryb and Th{\'{e}}bault}{Gryb and
  Th{\'{e}}bault}{2016a}]{Gryb2016b}
Gryb, S. and K.~P.~Y. Th{\'{e}}bault (2016a).
\newblock Regarding the `hole argument' and the `problem of time'.
\newblock {\em Philosophy of Science\/}~{\em 83\/}(4), 563--584.

\bibitem[\protect\citeauthoryear{Gryb and Th{\'{e}}bault}{Gryb and
  Th{\'{e}}bault}{2016b}]{Gryb:2016a}
Gryb, S. and K.~P.~Y. Th{\'{e}}bault (2016b).
\newblock {Schr{\"o}dinger Evolution for the Universe: Reparametrization}.
\newblock {\em Classical and Quantum Gravity\/}~{\em 33\/}(6), 065004.

\bibitem[\protect\citeauthoryear{Gryb and Th{\'{e}}bault}{Gryb and
  Th{\'{e}}bault}{2016c}]{gryb:2016c}
Gryb, S. and K.~P.~Y. Th{\'{e}}bault (2016c).
\newblock Time remains.
\newblock {\em The British Journal for the Philosophy of Science\/}~{\em 67},
  663--705.

\bibitem[\protect\citeauthoryear{Gryb and Th\'{e}bault}{Gryb and
  Th\'{e}bault}{2023}]{Gryb:2023}
Gryb, S. and K.~P.~Y. Th\'{e}bault (2023).
\newblock {\em Time Regained: Symmetry and Evolution in Classical Mechanics}.
\newblock Oxford University Press.

\bibitem[\protect\citeauthoryear{Hawking}{Hawking}{1969}]{hawking:1969}
Hawking, S.~W. (1969).
\newblock The existence of cosmic time functions.
\newblock {\em Proceedings of the Royal Society of London. Series A.
  Mathematical and Physical Sciences\/}~{\em 308\/}(1494), 433--435.

\bibitem[\protect\citeauthoryear{H{\"o}hn}{H{\"o}hn}{2019}]{hohn:2019b}
H{\"o}hn, P.~A. (2019, May).
\newblock Switching internal times and a new perspective on the `wave function
  of the universe'.
\newblock {\em Universe\/}~{\em 5\/}(5), 116.

\bibitem[\protect\citeauthoryear{H{\"o}hn, Smith, and Lock}{H{\"o}hn
  et~al.}{2021}]{hohn:2021}
H{\"o}hn, P.~A., A.~R. Smith, and M.~P. Lock (2021).
\newblock Trinity of relational quantum dynamics.
\newblock {\em Physical Review D\/}~{\em 104\/}(6), 066001.

\bibitem[\protect\citeauthoryear{Isham}{Isham}{1993}]{Isham:1992ms}
Isham, C.~J. (1993).
\newblock {Canonical quantum gravity and the problem of time}.
\newblock {\em NATO Sci. Ser. C\/}~{\em 409}, 157--287.

\bibitem[\protect\citeauthoryear{Kabel, de~la Hamette, Apadula, Cepollaro,
  Gomes, Butterfield, and Brukner}{Kabel et~al.}{2024}]{QuantumHole}
Kabel, V., A.-C. de~la Hamette, L.~Apadula, C.~Cepollaro, H.~Gomes,
  J.~Butterfield, and C.~Brukner (2024).
\newblock Identification is pointless: Quantum reference frames, localisation
  of events, and the quantum hole argument.

\bibitem[\protect\citeauthoryear{Kolb and Turner}{Kolb and
  Turner}{1994}]{kolb1994early}
Kolb, E. and M.~Turner (1994).
\newblock {\em The Early Universe}.
\newblock Frontiers in physics. Perseus.

\bibitem[\protect\citeauthoryear{Kuchar}{Kuchar}{1999}]{kuchar:1999}
Kuchar, K. (1999).
\newblock The problem of time in quantum geometrodynamics.
\newblock {\em The arguments of time\/}, 169--196.

\bibitem[\protect\citeauthoryear{Kucha\u{r}}{Kucha\u{r}}{1991}]{Kuchar:1991}
Kucha\u{r}, K. (1991).
\newblock The problem of time in canonical quantization of relativistic
  systems.
\newblock In A.~Ashtekar and J.~Stachel (Eds.), {\em Conceptual Problems of
  Quantum Gravity}, pp.\  141. Boston University Press.

\bibitem[\protect\citeauthoryear{Kucha\u{r}}{Kucha\u{r}}{1992}]{Kuchar:1992}
Kucha\u{r}, K. (1992).
\newblock Time and interpretations of quantum gravity.
\newblock In {\em 4th Canadian Conference on General Relativity {\ldots}}.
  World Scientific Singapore.

\bibitem[\protect\citeauthoryear{Landau and Lifshitz}{Landau and
  Lifshitz}{1987}]{Landau1987-fh}
Landau, L.~D. and E.~M. Lifshitz (1987, January).
\newblock {\em The classical theory of fields\/} (4 ed.).
\newblock Oxford, England: Butterworth-Heinemann.

\bibitem[\protect\citeauthoryear{Lema{\^\i}tre}{Lema{\^\i}tre}{1931}]{Lematre1931}
Lema{\^\i}tre, A.~G. (1931, March).
\newblock A homogeneous universe of constant mass and increasing radius
  accounting for the radial velocity of extra-galactic nebulae.
\newblock {\em Monthly Notices of the Royal Astronomical Society\/}~{\em
  91\/}(5), 483--490.

\bibitem[\protect\citeauthoryear{Malament}{Malament}{2012a}]{Malament:2012}
Malament, D.~B. (2012a).
\newblock {\em Topics in the foundations of general relativity and Newtonian
  gravitation theory}.
\newblock University of Chicago Press.

\bibitem[\protect\citeauthoryear{Malament}{Malament}{2012b}]{malament2012topics}
Malament, D.~B. (2012b).
\newblock {\em Topics in the foundations of general relativity and Newtonian
  gravitation theory}.
\newblock University of Chicago Press.

\bibitem[\protect\citeauthoryear{Misner, Thorne, Wheeler, and Kaiser}{Misner
  et~al.}{2017}]{misner2017gravitation}
Misner, C., K.~Thorne, J.~Wheeler, and D.~Kaiser (2017).
\newblock {\em Gravitation}.
\newblock Princeton University Press.

\bibitem[\protect\citeauthoryear{Norton}{Norton}{1989}]{Norton1989}
Norton, J. (1989, October).
\newblock Coordinates and covariance: Einstein{\textquotesingle}s view of
  space-time and the modern view.
\newblock {\em Foundations of Physics\/}~{\em 19\/}(10), 1215--1263.

\bibitem[\protect\citeauthoryear{Norton}{Norton}{1993}]{Norton1993-NORGCA}
Norton, J.~D. (1993).
\newblock General covariance and the foundations of general relativity: Eight
  decades of dispute.
\newblock {\em Reports of Progress in Physics\/}~{\em 56}, 791--861.

\bibitem[\protect\citeauthoryear{Page and Wootters}{Page and
  Wootters}{1983}]{page:1983}
Page, D.~N. and W.~K. Wootters (1983).
\newblock Evolution without evolution: Dynamics described by stationary
  observables.
\newblock {\em Physical Review D\/}~{\em 27\/}(12), 2885.

\bibitem[\protect\citeauthoryear{Penrose}{Penrose}{1980}]{Penrose:1980ge}
Penrose, R. (1980).
\newblock {\em {Singularities and Time Asymmetry}}, pp.\  581--638.

\bibitem[\protect\citeauthoryear{Perivolaropoulos and Skara}{Perivolaropoulos
  and Skara}{2022}]{perivolaropoulos:2022}
Perivolaropoulos, L. and F.~Skara (2022).
\newblock Challenges for $\lambda$cdm: An update.
\newblock {\em New Astronomy Reviews\/}~{\em 95}, 101659.

\bibitem[\protect\citeauthoryear{Pitts}{Pitts}{2019}]{pitts:2019}
Pitts, J.~B. (2019).
\newblock What are observables in hamiltonian einstein--maxwell theory?
\newblock {\em Foundations of Physics\/}, 1--11.

\bibitem[\protect\citeauthoryear{Pooley}{Pooley}{2017}]{pooley:2017}
Pooley, O. (2017).
\newblock Background independence, diffeomorphism invariance and the meaning of
  coordinates.
\newblock {\em Towards a theory of spacetime theories\/}, 105--143.

\bibitem[\protect\citeauthoryear{Read}{Read}{2023}]{James_Read2023-mk}
Read, J. (2023, November).
\newblock {\em Background independence in classical and quantum gravity}.
\newblock London, England: Oxford University Press.

\bibitem[\protect\citeauthoryear{Reichenbach}{Reichenbach}{1928}]{reichenbach:1928}
Reichenbach, H. (1928).
\newblock {\em The philosophy of space and time}.
\newblock Berlin:: Walter de Gruyter.

\bibitem[\protect\citeauthoryear{Rickles}{Rickles}{2007}]{Rickles:2007}
Rickles, D. (2007).
\newblock {\em Symmetry, Structure, and Spacetime}, Volume~3 of {\em Philossphy
  and Foundations of Physics}.
\newblock Elsevier.

\bibitem[\protect\citeauthoryear{Robertson}{Robertson}{1935}]{Robertson1935}
Robertson, H.~P. (1935, November).
\newblock Kinematics and world-structure.
\newblock {\em The Astrophysical Journal\/}~{\em 82}, 284.

\bibitem[\protect\citeauthoryear{Roser and Valentini}{Roser and
  Valentini}{2014}]{Roser2014}
Roser, P. and A.~Valentini (2014, November).
\newblock Classical and quantum cosmology with york time.
\newblock {\em Classical and Quantum Gravity\/}~{\em 31\/}(24), 245001.

\bibitem[\protect\citeauthoryear{Rovelli}{Rovelli}{1991a}]{rovelli1991a}
Rovelli, C. (1991a).
\newblock Time in quantum gravity: An hypothesis.
\newblock {\em Physical Review D\/}~{\em 43\/}(2), 442.

\bibitem[\protect\citeauthoryear{Rovelli}{Rovelli}{1991b}]{Rovelli1991}
Rovelli, C. (1991b, February).
\newblock What is observable in classical and quantum gravity?
\newblock {\em Classical and Quantum Gravity\/}~{\em 8\/}(2), 297--316.

\bibitem[\protect\citeauthoryear{Rovelli}{Rovelli}{2002a}]{RovelliGPS}
Rovelli, C. (2002a, Jan).
\newblock Gps observables in general relativity.
\newblock {\em Phys. Rev. D\/}~{\em 65}, 044017.

\bibitem[\protect\citeauthoryear{Rovelli}{Rovelli}{2002b}]{Rovelli_2002}
Rovelli, C. (2002b, jun).
\newblock Partial observables.
\newblock {\em Physical Review D\/}~{\em 65\/}(12).

\bibitem[\protect\citeauthoryear{Rovelli}{Rovelli}{2004}]{Rovelli2004}
Rovelli, C. (2004).
\newblock {\em Quantum Gravity}.
\newblock Cambridge University Press.

\bibitem[\protect\citeauthoryear{Rovelli}{Rovelli}{2007}]{Rovelli:2007}
Rovelli, C. (2007).
\newblock Comment on "are the spectra of geometrical operators in loop quantum
  gravity really discrete?" by b. dittrich and t. thiemann.

\bibitem[\protect\citeauthoryear{Rovelli}{Rovelli}{2014}]{Rovelli2014}
Rovelli, C. (2014, January).
\newblock Why gauge?
\newblock {\em Foundations of Physics\/}~{\em 44\/}(1), 91--104.

\bibitem[\protect\citeauthoryear{Rovelli}{Rovelli}{2024}]{RovelliPrinceton}
Rovelli, C. (2024).
\newblock Princeton seminars on physics and philosophy.

\bibitem[\protect\citeauthoryear{Rugh and Zinkernagel}{Rugh and
  Zinkernagel}{2009}]{rugh:2009}
Rugh, S.~E. and H.~Zinkernagel (2009).
\newblock On the physical basis of cosmic time.
\newblock {\em Studies in History and Philosophy of Science Part B: Studies in
  History and Philosophy of Modern Physics\/}~{\em 40\/}(1), 1--19.

\bibitem[\protect\citeauthoryear{Ryden}{Ryden}{2016}]{Ryden_2016}
Ryden, B. (2016).
\newblock {\em Newton versus Einstein}, pp.\  27--48.
\newblock Cambridge University Press.

\bibitem[\protect\citeauthoryear{Schander and Thiemann}{Schander and
  Thiemann}{2021}]{Schander2021}
Schander, S. and T.~Thiemann (2021, July).
\newblock Backreaction in cosmology.
\newblock {\em Frontiers in Astronomy and Space Sciences\/}~{\em 8}.

\bibitem[\protect\citeauthoryear{Smeenk}{Smeenk}{2013}]{smeenk:2013}
Smeenk, C. (2013).
\newblock Time in cosmology.
\newblock {\em A Companion to the Philosophy of Time\/}, 201--219.

\bibitem[\protect\citeauthoryear{Smeenk}{Smeenk}{2022}]{Smeenk_2022}
Smeenk, C. (2022).
\newblock Trouble with hubble: Status of the big bang models.
\newblock {\em Philosophy of Science\/}~{\em 89\/}(5), 1265--1274.

\bibitem[\protect\citeauthoryear{Tal}{Tal}{2016}]{tal:2016}
Tal, E. (2016).
\newblock Making time: A study in the epistemology of measurement.
\newblock {\em The British Journal for the Philosophy of Science\/}~{\em
  67\/}(1), 297--335.

\bibitem[\protect\citeauthoryear{Tambornino}{Tambornino}{2012}]{Tambornino2012}
Tambornino, J. (2012, March).
\newblock Relational observables in gravity: a review.
\newblock {\em Symmetry, Integrability and Geometry: Methods and
  Applications\/}.

\bibitem[\protect\citeauthoryear{Teitelboim}{Teitelboim}{1973}]{Teitelboim:1973}
Teitelboim, C. (1973).
\newblock How commutators of constraints reflect the spacetime structure.
\newblock {\em Annals of Physics\/}~{\em 79\/}(2), 542--557.

\bibitem[\protect\citeauthoryear{Th{\'e}bault}{Th{\'e}bault}{2021a}]{thebault:2021b}
Th{\'e}bault, K.~P. (2021a).
\newblock On mach on time.
\newblock {\em Studies in History and Philosophy of Science Part A\/}~{\em 89},
  84--102.

\bibitem[\protect\citeauthoryear{Th{\'e}bault}{Th{\'e}bault}{2021b}]{thebault:2019}
Th{\'e}bault, K.~P. (2021b).
\newblock The problem of time.
\newblock In A.~Wilson and E.~Knox (Eds.), {\em Routledge Companion to
  Philosophy of Physics}. Routledge.

\bibitem[\protect\citeauthoryear{Thiemann}{Thiemann}{2006}]{Thiemann-k-essence}
Thiemann, T. (2006).
\newblock Solving the problem of time in general relativity and cosmology with
  phantoms and k -- essence.

\bibitem[\protect\citeauthoryear{Thiemann}{Thiemann}{2007}]{Thiemann:2007}
Thiemann, T. (2007).
\newblock {\em Modern canonical quantum general relativity}.
\newblock Cambridge University Press.

\bibitem[\protect\citeauthoryear{Turner}{Turner}{2000}]{TURNER2000619}
Turner, M.~S. (2000).
\newblock The dark side of the universe: from zwicky to accelerated expansion.
\newblock {\em Physics Reports\/}~{\em 333-334}, 619--635.

\bibitem[\protect\citeauthoryear{Wald}{Wald}{1984}]{Wald1984}
Wald, R.~M. (1984, June).
\newblock {\em General Relativity}.
\newblock Chicago, IL: University of Chicago Press.

\bibitem[\protect\citeauthoryear{Walker}{Walker}{1937}]{Walker1937}
Walker, A.~G. (1937).
\newblock On milne's theory of world-structure*.
\newblock {\em Proceedings of the London Mathematical Society\/}~{\em
  s2-42\/}(1), 90--127.

\bibitem[\protect\citeauthoryear{Wallace}{Wallace}{2024}]{WALLACE202438}
Wallace, D. (2024).
\newblock Gauge invariance through gauge fixing.
\newblock {\em Studies in History and Philosophy of Science\/}~{\em 108},
  38--45.

\bibitem[\protect\citeauthoryear{Weinberg}{Weinberg}{1972}]{weinberg1972gravitation}
Weinberg, S. (1972).
\newblock {\em Gravitation and Cosmology: Principles and Applications of the
  General Theory of Relativity}.
\newblock Wiley.

\bibitem[\protect\citeauthoryear{York}{York}{1972}]{york:1972}
York, J.~W. (1972).
\newblock Role of conformal three-geometry in the dynamics of gravitation.
\newblock {\em Physical review letters\/}~{\em 28\/}(16), 1082.

\bibitem[\protect\citeauthoryear{York}{York}{1973}]{york:1973}
York, J.~W. (1973).
\newblock Conformally invariant orthogonal decomposition of symmetric tensors
  on riemannian manifolds and the initial-value problem of general relativity.
\newblock {\em Journal of Mathematical Physics\/}~{\em 14\/}(4), 456--464.

\end{thebibliography}

\appendix
\section{Gauge Freedom and Gauge Fixings}
\label{gauge}

The framework of Dirac observables has as its intended goal the removal of underdetermination in phase space dynamics due to the presence of gauge freedom. This is clearly the important point and not whether or not the letter of the Dirac criterion or the spirit of general covariance have been respected. 
The observables of the theory are required to be gauge-invariant quantities. Here we briefly summarise some formal aspects relating gauge invariance and gauge fixings in order to clarify the formal status of cosmic proper time within the foundations of general relativity. See \cite{Gomes2024} for further details.

In the context of general relativity the desire is to formulate the representations of observables such that they are invariant under diffeomorphisms. A space of models $\Phi$ within GR can be seen as a principal bundle with $Diff(\mathcal{M})$ as its structure group and $[\Phi]:=\{[\varphi], \varphi\in \Phi)\}$ its base space.  Selecting a reference frame amounts to defineing a \textit{unique} section map $\sigma:[\varphi]\rightarrow \sigma([\varphi])\in \Phi$, where the choice of a section is a smooth injection from the space of equivalence classes of models to the space of models, and corresponds to a choice of a submanifold on the fibre-bundle that intersect each fibre $\mathbb{F}_\varphi:=\textbf{pr}^{-1}([\varphi])$ exactly once (with $\textbf{pr}:\varphi \rightarrow [\varphi]$).

\begin{figure}[!h]
    \centering
    \includegraphics[scale=0.42]{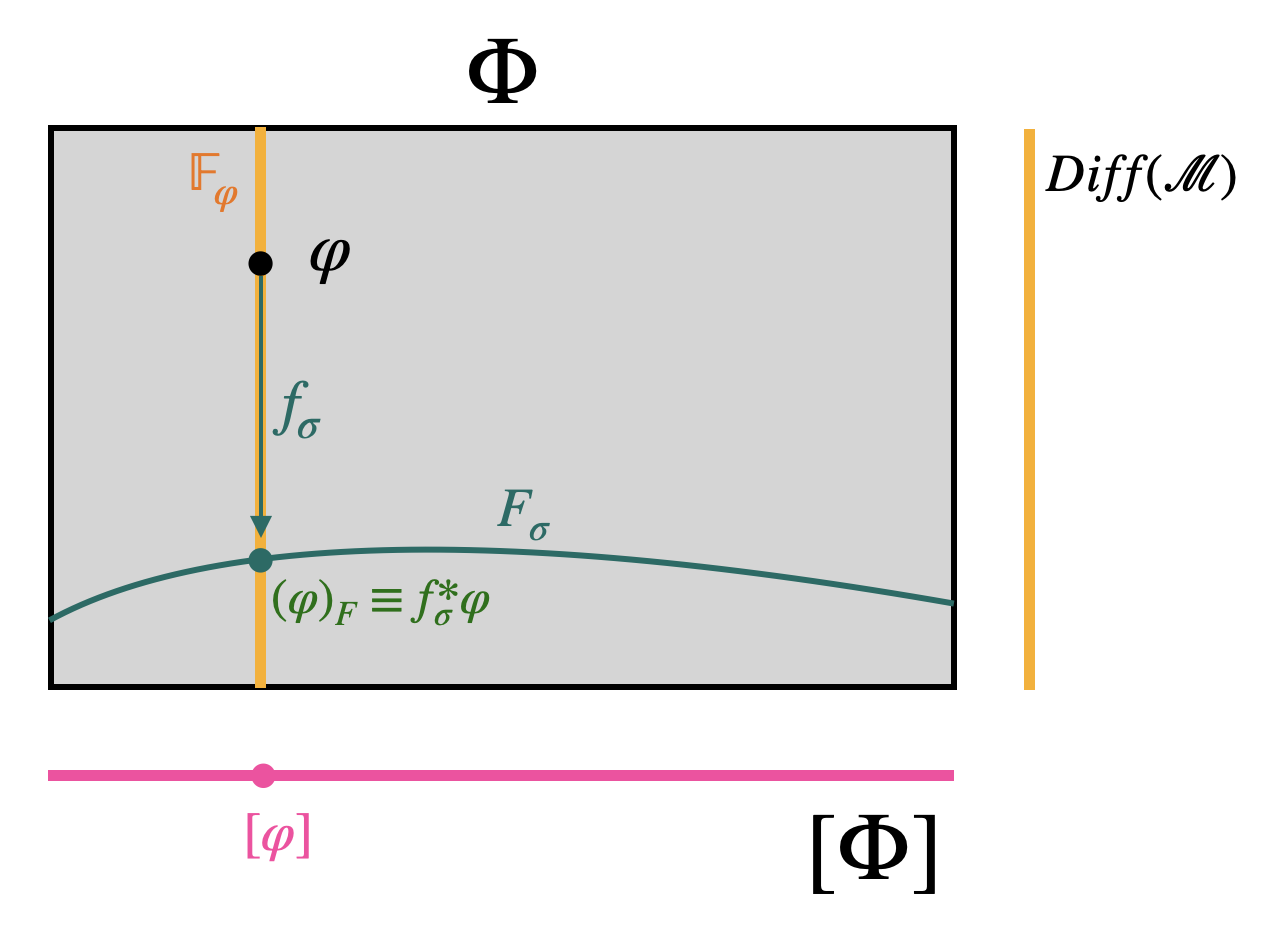}
    \caption{The space of models $\Phi$. Each point corresponds to a particular configuration $\varphi$. A reference frame $\sigma$ picks out a \textit{unique} representative $(\varphi)_F$ for each fibre $\mathbb{F}_\varphi$. Models belonging to the same fibre are taken to be physically equivalent, since a fibre corresponds to a gauge orbit.}
    \label{fiberbundle}
\end{figure}

Equivalently we can make use of the so-called projection operator $f_{\sigma}:\varphi \rightarrow f_{\sigma}^*\varphi$, an equivalent of the section map, which takes any element of a given fibre to the \textit{unique} image of the section. It is an embedding map, \textit{acting within a fibre} and it is characterised by the auxiliary condition $F_{\sigma}(\varphi)=0$, making the choice of a reference frame (or a section) analogous to a gauge-fixing procedure.  The use of the projection operator $f_\sigma:\Phi\rightarrow \Phi$ instead of the section map $\sigma:[\Phi]\rightarrow \Phi$ codifies the \textit{symmetry-first} or \textit{external sophistication} principle that it is unnecessary to \textit{intrinsically} represente elements $[\varphi]$ of $[\Phi]$ through a parametrisation of $[\Phi]$. Thus it rejects the \textit{structure-first, or internal approach} \citep{Dewar2019-DEWSAS-4,jacobs2021symmetries}. The quantity resulting from the choice of a section is the relational, gauge-invariant observable $f_\sigma^*\varphi \equiv (\varphi)_F$. See Figure \ref{fiberbundle}.

The important point is that the transformation that changes the reference frame corresponds to a \textit{change of section}. So, it should \textit{not} be understood as something that acts on the fields configuration: it does not act on the dynamically possible models $\varphi$, but acts \textit{directly} on the already constructed gauge-invariant observables, changing frames (section) and getting us to a different and \textit{new} observable, i.e. a new representative of a fibre. This substantiates Thiemann's claim that a change of reference frame has observable consequences for the dynamics (see section \ref{sec5}).

In order for the cosmic proper time $\tau$ to be considered a physical clock dynamically coupled with the Hubble parameter (and thus with the metric), the following condition must be met: If $(g_{ab},\tau)$ is a dynamically possible model of the theory, then neither $([d^*g]_{ab},\tau)$ nor $(g_{ab},d^*\tau)$ is, $\forall d \in Diff(\mathcal{M})$. Thus, the choice of $\tau$ as the \textbf{RRF} clock (rather than one of its diffeomorphic copies) provide a \textit{unique} representation of $H(\tau):= \big[\tau^{-1}\big]^* H$ for some initial data, which is thus a \textit{bona-fide} gauge-invariant, complete observable.\footnote{With the symbol $\big[ \bullet \big]^*$, we denote the pullback.} For this reason, $\tau$ fixes the gauge for the FLRW metric and by definition is such that there is no longer gauge freedom in the theory, and the potential for underdetermination.  However, given the correspondence between the choice of a reference frame and the choice of a gauge \citep{Bamonti2023,Gomes2024}, we always have the possibility of changing reference frames and obtaining new relational observables.

\end{document}